\input harvmac
\let\includefigures=\iftrue
\let\useblackboard=\iftrue
\newfam\black

\includefigures
\message{If you do not have epsf.tex (to include figures),}
\message{change the option at the top of the tex file.}
\input epsf
\def\figin{\epsfcheck\figin}\def\figins{\epsfcheck\figins}
\def\epsfcheck{\ifx\epsfbox\UnDeFiNeD
\message{(NO epsf.tex, FIGURES WILL BE IGNORED)}
\gdef\figin##1{\vskip2in}\gdef\figins##1{\hskip.5in}
\else\message{(FIGURES WILL BE INCLUDED)}%
\gdef\figin##1{##1}\gdef\figins##1{##1}\fi}
\def\DefWarn#1{}
\def\figinsert{\goodbreak\midinsert}
\def\ifig#1#2#3{\DefWarn#1\xdef#1{fig.~\the\figno}
\writedef{#1\leftbracket fig.\noexpand~\the\figno}%
\figinsert\figin{\centerline{#3}}\medskip\centerline{\vbox{
\baselineskip12pt\advance\hsize by -1truein
\noindent\footnotefont{\bf Fig.~\the\figno:} #2}}
\endinsert\global\advance\figno by1}
\else
\def\ifig#1#2#3{\xdef#1{fig.~\the\figno}
\writedef{#1\leftbracket fig.\noexpand~\the\figno}%
\global\advance\figno by1} \fi

\def\id{{1 \kern-.28em {\rm l}}}

\def\K3{{\bf K3}}
\def\journal#1&#2(#3){\unskip, \sl #1\ \bf #2 \rm(19#3) }
\def\andjournal#1&#2(#3){\sl #1~\bf #2 \rm (19#3) }

\def\bar{\overline}

\def\ie{{\it i.e.}}
\def\eg{{\it e.g.}}

\def\tilde{\widetilde}

\def\frac#1#2{{#1\over#2}}

\def\half{\frac12}

\def\inbar{\,\vrule height1.5ex width.4pt depth0pt}
\def\IC{\relax\hbox{$\inbar\kern-.3em{\rm C}$}}
\def\IR{\relax{\rm I\kern-.18em R}}
\def\IP{\relax{\rm I\kern-.18em P}}

%
%

%
\catcode`\@=11
\def\slash#1{\mathord{\mathpalette\c@ncel{#1}}}
\overfullrule=0pt

\def\underrel#1\over#2{\mathrel{\mathop{\kern\z@#1}\limits_{#2}}}

\catcode`\@=12


%

\def\exp{{\rm exp}}


\lref\CallanAT{
C.~G.~Callan, J.~A.~Harvey and A.~Strominger,
``Supersymmetric string solitons,''
arXiv:hep-th/9112030.
}

\lref\khu{ R.~R.~Khuri, ``Remark on string solitons,'' Phys.\
Rev.\ D {\bf 48}, 2947 (1993) [arXiv:hep-th/9305143].
}

\lref\ghm{ M.~B.~Green, J.~A.~Harvey and G.~W.~Moore, ``I-brane
inflow and anomalous couplings on D-branes,'' Class.\ Quant.\
Grav.\  {\bf 14}, 47 (1997) [arXiv:hep-th/9605033].
}

\lref\pw{ A.~M.~Polyakov and P.~B.~Wiegmann, ``Theory Of
Nonabelian Goldstone Bosons In Two Dimensions,'' Phys.\ Lett.\ B
{\bf 131}, 121 (1983).
}

\lref\ElitzurMM{ S.~Elitzur, O.~Feinerman, A.~Giveon and
D.~Tsabar, ``String theory on $AdS(3) \times S(3) \times S(3) \times S(1)$,''
Phys.\ Lett.\ B {\bf 449}, 180 (1999) [arXiv:hep-th/9811245].
}

\lref\AharonyUB{
O.~Aharony, M.~Berkooz, D.~Kutasov and N.~Seiberg,
``Linear dilatons, NS5-branes and holography,''
JHEP {\bf 9810}, 004 (1998)
[arXiv:hep-th/9808149].
}

\lref\GiveonPX{
A.~Giveon and D.~Kutasov,
``Little string theory in a double scaling limit,''
JHEP {\bf 9910}, 034 (1999)
[arXiv:hep-th/9909110].
}

\lref\GiveonTQ{
A.~Giveon and D.~Kutasov,
``Comments on double scaled little string theory,''
JHEP {\bf 0001}, 023 (2000)
[arXiv:hep-th/9911039].
}

\lref\BerkoozCQ{
M.~Berkooz, M.~Rozali and N.~Seiberg,
``Matrix description of M theory on $T^4$ and $T^5$,''
Phys.\ Lett.\ B {\bf 408}, 105 (1997)
[arXiv:hep-th/9704089].
}

\lref\SeibergZK{
N.~Seiberg,
``New theories in six dimensions and matrix description of M-theory on  $T^5$
and $T^5/Z(2)$,''
Phys.\ Lett.\ B {\bf 408}, 98 (1997)
[arXiv:hep-th/9705221].
}

\lref\AharonyXN{
O.~Aharony, A.~Giveon and D.~Kutasov,
``LSZ in LST,''
Nucl.\ Phys.\ B {\bf 691}, 3 (2004)
[arXiv:hep-th/0404016].
}

\lref\AharonyVK{
O.~Aharony, B.~Fiol, D.~Kutasov and D.~A.~Sahakyan,
``Little string theory and heterotic/type II duality,''
Nucl.\ Phys.\ B {\bf 679}, 3 (2004)
[arXiv:hep-th/0310197].
}

\lref\MaldacenaCG{ J.~M.~Maldacena and A.~Strominger,
``Semiclassical decay of near-extremal fivebranes,'' JHEP {\bf
9712}, 008 (1997) [arXiv:hep-th/9710014].
}

\lref\GukovYM{ S.~Gukov, E.~Martinec, G.~W.~Moore and
A.~Strominger, ``The search for a holographic dual to $AdS(3)\times
S^3 \times S^3 \times S^1$,'' arXiv:hep-th/0403090.
}

\lref\deBoerRH{
J.~de Boer, A.~Pasquinucci and K.~Skenderis,
``AdS/CFT dualities involving large 2d N = 4 superconformal symmetry,''
Adv.\ Theor.\ Math.\ Phys.\  {\bf 3}, 577 (1999)
[arXiv:hep-th/9904073].
}

\lref\KutasovUA{
D.~Kutasov and N.~Seiberg,
``Noncritical Superstrings,''
Phys.\ Lett.\ B {\bf 251}, 67 (1990).
}

\lref\ItzhakiDD{ N.~Itzhaki, J.~M.~Maldacena, J.~Sonnenschein and
S.~Yankielowicz, ``Supergravity and the large N limit of theories
with sixteen  supercharges,'' Phys.\ Rev.\ D {\bf 58}, 046004
(1998) [arXiv:hep-th/9802042].
}

\lref\KarczmarekBW{
J.~L.~Karczmarek, J.~Maldacena and A.~Strominger,
``Black hole non-formation in the matrix model,''
arXiv:hep-th/0411174.
}

\lref\GiveonMI{
A.~Giveon, D.~Kutasov, E.~Rabinovici and A.~Sever,
``Phases of quantum gravity in AdS(3) and linear dilaton backgrounds,''
Nucl.\ Phys.\ B {\bf 719}, 3 (2005)
[arXiv:hep-th/0503121].
      }

\lref\GiveonZM{ A.~Giveon, D.~Kutasov and O.~Pelc, ``Holography
for non-critical superstrings,'' JHEP {\bf 9910}, 035 (1999)
[arXiv:hep-th/9907178].
}

\lref\maldanew{J.~Maldacena and H. Lin, to appear; J. Maldacena,
talk at Strings 2005. }

\lref\SchwarzYJ{ J.~H.~Schwarz, ``Superconformal Chern-Simons
theories,'' JHEP {\bf 0411}, 078 (2004) [arXiv:hep-th/0411077].
}

\lref\GreeneHU{
B.~R.~Greene, D.~R.~Morrison and A.~Strominger,
``Black hole condensation and the unification of string vacua,''
Nucl.\ Phys.\ B {\bf 451}, 109 (1995)
[arXiv:hep-th/9504145].
}

\Title{\vbox{\baselineskip12pt\hbox{hep-th/0508025}
\hbox{PUPT/2168}}} {\vbox{\centerline{I-Brane Dynamics }}}
\bigskip

\centerline{\it Nissan Itzhaki $ ^{a,b}$, David Kutasov $ ^c$ and
Nathan Seiberg $ ^b$}
\bigskip
\centerline{$ ^a$Department of Physics, Princeton University,
Princeton, NJ 08544}
\medskip
\centerline{$ ^b$School of Natural Sciences, Institute for
Advanced Study} \centerline{Einstein Drive, Princeton, NJ 08540}
\medskip
\centerline{$ ^c$EFI and Department of Physics, University of
Chicago}\centerline{5640 S. Ellis Av. Chicago, IL 60637}

\smallskip

\vglue .3cm

\bigskip

\bigskip
 \noindent
We study the dynamics near a $1+1$ dimensional intersection of two
orthogonal stacks of fivebranes in type IIB string theory, using
an open string description valid at weak coupling, and a closed
string description valid at strong coupling. The weak coupling
description suggests that this system is invariant under eight
supercharges with a particular chirality in $1+1$ dimensions, and
its spectrum contains chiral fermions localized at the intersection.
The closed string description leads to a rather different picture --
a three dimensional Poincare invariant theory with a gap and sixteen
supercharges. We show that this dramatic change in the behavior of
the system is partly due to anomaly inflow. Taking it into account
leads to a coherent picture, both when the fivebranes in each stack
are coincident and when they are separated.

\bigskip

\Date{August 2005}

\noindent

\newsec{Introduction and summary}

In this paper we  study a system consisting of two stacks of
fivebranes in type IIB string theory, which intersect on an
$\IR^{1,1}$. This system can be analyzed from a number of
different points of view and exhibits some interesting features:

\item{(1)} Holography: one might expect that the dynamics at the
$1+1$ dimensional intersection of the two sets of fivebranes should
be holographically related to a $2+1$ dimensional bulk theory,
with the extra dimension being the radial direction away from the
intersection. In fact the bulk  description includes {\it two} radial
directions away from each set of fivebranes, and is $3+1$ dimensional.
The corresponding boundary theory is $2+1$ dimensional.

\item{(2)} Near-horizon symmetry enhancement: inspection of the
brane configuration suggests that the theory at the intersection
of the fivebranes should be invariant under $1+1$  dimensional
Poincare symmetry,  $ISO(1,1)$, and eight supercharges with a
particular chirality in $1+1$ dimensions. However, the
near-horizon geometry describes a $2+1$ dimensional theory with
Poincare symmetry $ISO(2,1)$, and twice as many supercharges as
one would naively expect -- eight copies of the two dimensional
spinor of ${\rm Spin}(2,1)$.  Interestingly, the superalgebra is
not the standard ${\cal N}=8$ supersymmetry in three
dimenions.\foot{This point was independently noticed in
\maldanew.}  This symmetry enhancement is surprising since
it implies that there are no normalizable states localized at
the intersection, while the weakly coupled open string analysis
leads to a large number of such states. In
particular, the intersection is expected to carry chiral degrees
of freedom in $1+1$ dimensions, which cannot be lifted, and whose
presence is inconsistent with the enhanced (super-) symmetry of the
near-horizon geometry.

\item{(3)} Anomaly inflow: the chiral modes living at the
intersection of the fivebranes carry non-zero anomalies under the
gauge symmetries on the fivebranes. These anomalies are cancelled
by inflow from the bulk of the fivebranes \ghm. The fact that the
anomaly inflow mechanism is at play here is the first sign that
the dynamics at the intersection is not decoupled from that in the
bulk of the fivebranes. As we will see, the higher dimensional
perspective is essential in understanding various features of this
system.

\subsec{The fivebrane system}

A concrete realization of our brane system is the following.
Consider $k_1$ $D5$-branes stretched in the directions $(012345)$,
and $k_2$ $D5$-branes stretched in $(016789)$. This configuration
is invariant under $ISO(1,1)\times SO(4)_{2345}\times
SO(4)_{6789}$. It also preserves eight supercharges, which satisfy
the constraints
 \eqn\conssusy{\eqalign{ \epsilon_L=&\Gamma^{012345}\epsilon_R , \cr
 \epsilon_L=&\Gamma^{016789}\epsilon_R~.\cr }}
Taking into account the fact that the spinors $\epsilon_L$,
$\epsilon_R$ in \conssusy\ have the same chirality in $9+1$
dimensions, one finds that the solutions of \conssusy\ are chiral
in $\IR^{1,1}$.

At long distances, one can describe the dynamics of this brane
configuration by a low energy effective field theory. In addition
to the familiar six dimensional $U(k_i)$ gauge theory with sixteen
supercharges on each set of $D5$-branes, the low lying spectrum
contains $k_1k_2$ complex, chiral (Weyl) fermions, which transform
in the representation $({\bf k_1, \bar{k}_2})$ of the two gauge
groups. These fermions originate  from fundamental strings
stretched between the two stacks of fivebranes. We will take them to
be left-moving below. Their presence breaks the
left-moving\foot{This should not be confused with the notion of
left and right in equation \conssusy, which refers to worldsheet
chirality.} spacetime SUSY; the eight right-moving supercharges
remain unbroken.

The fermions have central charge $c_L=k_1k_2$, $c_R=0$, and thus
carry a non-zero gravitational anomaly in $1+1$ dimensions. As we
change various continuous parameters defining the brane
configuration, such as the string coupling and the separations
between different parallel fivebranes in a given stack,
$c_L-c_R=k_1k_2$ cannot change.

The fermions are
also anomalous under $SU(k_1)\times SU(k_2)\times U(1)$. To
exhibit the anomaly, it is convenient to use the fact that we
can replace them by the holomorphic current algebra
 \eqn\curralg{SU(k_1)_{k_2}\times SU(k_2)_{k_1}\times U(1)~.}
Hence, the $SU(k_1)$ gauge group on the first set of fivebranes
sees an anomaly proportional to $k_2$ localized on a codimension
four defect, and vice-versa. There is also a $U(1)$ anomaly coming
from the last factor in \curralg.

\subsec{Weak coupling analysis}

To study the dynamics of the fermions, it is important to include
in the effective action the terms that couple them to the six
dimensional gauge fields on the fivebranes and cancel the
anomalies. As we will see in sections 2 and 3, these have an
interesting dynamical effect. Naively, the fermions are localized
at the intersection point of the two sets of fivebranes.  In fact,
we will see that they are supported away from the intersection.
The amount by which they are displaced is governed by the six
dimensional gauge couplings of the $SU(k_i)$ gauge theories,
$g_i^2=g_s l_s^2$. We will see that each of the three factors in
\curralg\ is displaced in a different way in the transverse space.

The displacements, which are proportional to the gauge couplings
$g_i$, vanish in the weak coupling limit $g_s\to 0$. In this
limit, the picture becomes indistinguishable from what one finds
by studying open strings stretched between $D5$-branes. This is
not surprising, since in this limit all couplings, including those
between the modes at the intersection and those in the bulk of the
fivebranes, vanish. On the other hand, as $g_s$ increases, the
effect of the anomaly inflow becomes more pronounced, and the
picture deviates more from the weakly coupled open string one.

Let us describe the low energy field theory in more detail. Here
we discuss the case when all fivebranes in a given stack are
coincident. The case of separated fivebranes will be studied
below. We parameterize the $1+1$ dimensional intersection of the
two stacks of branes by $(x^0, x^1)$. The other eight coordinates
can be organized into two $\IR^4$'s which we will parameterize by
spherical coordinates. We will denote by $u$ the radial direction
along the first set of branes (it is the radial direction of the
$\IR^4$ transverse to the second set of branes); $v$ is the radial
direction along the second set of branes (and the radial direction
of the space transverse to the first set of branes). The
intersection is at $u=v=0$.

We will find it convenient to reduce the system on the two $S^3$'s
of these spherical coordinates and study the resulting effective three
dimensional theory.  In addition to the two coordinates $(x^0, x^1)$
the base space has another spatial direction made out of the
half-lines\foot{Recall that these are radial directions, therefore
$u, v \ge 0$.} labeled by $u$ and $v$, glued at $u=v=0$ (see figure 1).

\ifig\tspace{The spatial part of the $2+1$ dimensional space we
consider. One spatial direction is $x^1$ and the other is
constructed out of $u$ and $v$. Figure 1(a) stresses that in
the embedding space, the $u$ and $v$ directions are orthogonal.
Figure 1(b) stresses that $u$ and $v$ can be combined into a
second spatial direction.} {\epsfxsize5.0in\epsfbox{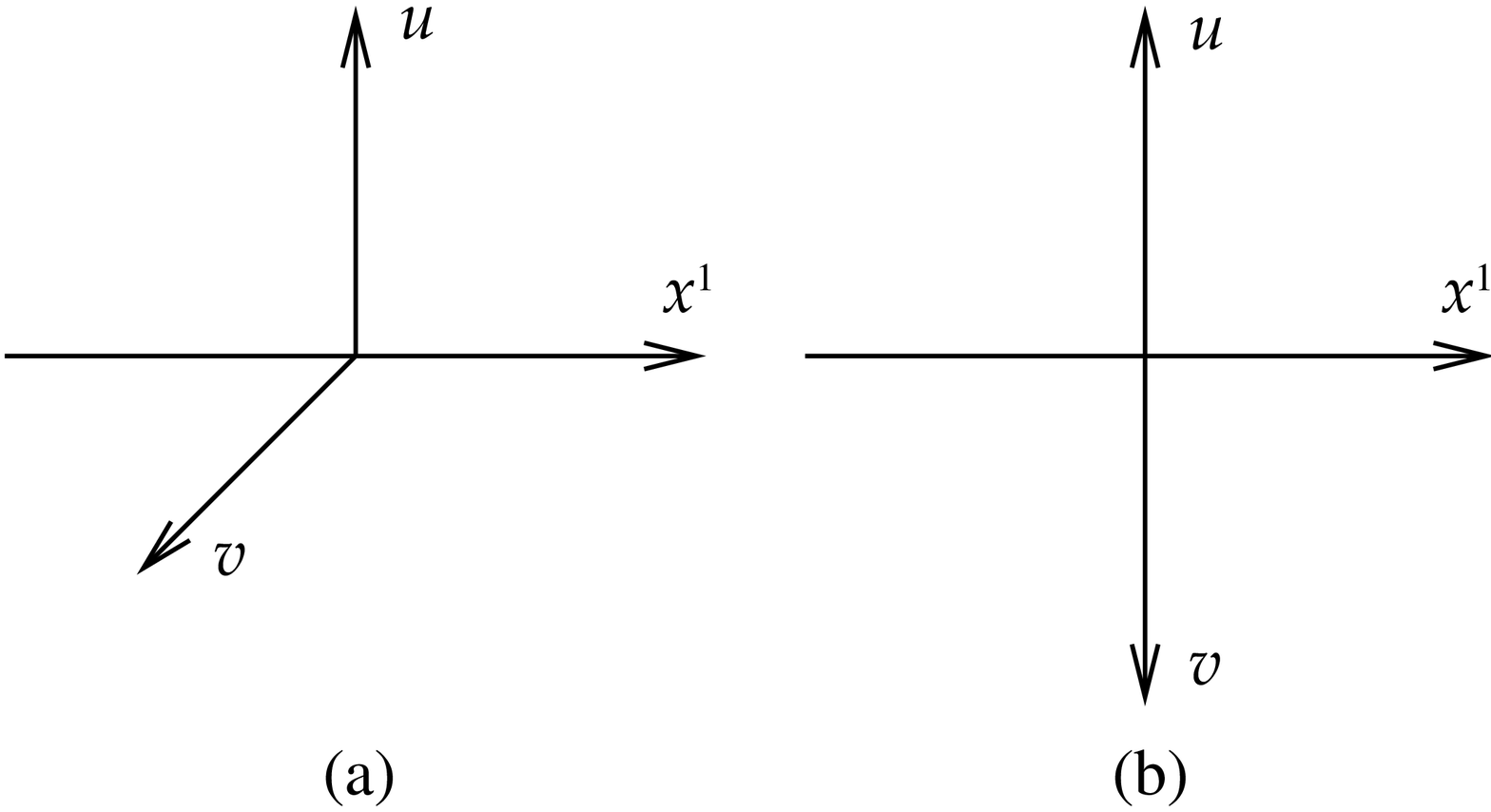}}

Thus we have a $2+1$ dimensional system with a ``domain
wall'' located on the real axis $u=v=0$. The coordinates along the
wall are $(x^0,x^1)$. For zero coupling the free fermions and their
current algebra \curralg\ are supported on the ``domain wall.''  In
addition to these fermions the system also has gauge fields which
propagate in the entire 2+1 dimensional space.  On the upper half
plane we have an $SU(k_1)$ gauge theory, while on the lower one we
have an $SU(k_2)$ gauge theory. There is also a $U(1)$ gauge field
which propagates on the whole plane. These theories have Chern-Simons
terms which are due to the fact that we are studying fivebranes
wrapped on three-spheres with non-vanishing three-form flux.
Because of these Chern-Simons terms the gauge fields are massive.

In figure 2 we describe the location of the chiral modes in our
$2+1$ dimensional system. Here we describe only the nontrivial
directions $u$ and $v$ and omit the $\IR^{1,1}$ labeled by
$(x^0, x^1)$.  In order to stress that $u$ and $v$ originate
from different branes,
and are orthogonal in the original $\IR^{9,1}$, the $u$ axis is
drawn orthogonal to the $v$ axis, as in fig. 1(a).  For zero
coupling the fermions are supported at $u=v=0$ (see figure 2(a)).
However, as we will see in sections 2 and 3, because of the
interaction with the gauge fields, when the gauge coupling
constants are nonzero, the current algebra is displaced. Different
parts of \curralg\ ``move'' to different locations (see figure 2(b)).
$SU(k_1)_{k_2}$ is supported at $u \sim \sqrt{g_1^2 k_2}$,
$SU(k_2)_{k_1}$ is supported at $v \sim \sqrt{g_2^2 k_1}$
and the $U(1)$ factor is supported at both these points.

\ifig\loc{(a) At zero coupling all the fermions are supported at
the intersection. (b) At finite coupling they move away from the
intersection.  Different parts of the current algebra move in
different directions.  $SU(k_1)_{k_2}$ is supported at $u \sim
\sqrt{g_1^2 k_2}$; $SU(k_2)_{k_1}$ is supported at $v \sim
\sqrt{g_2^2 k_1}$;  the $U(1)$ mode is supported at both
of these points.
} {\epsfxsize5.0in\epsfbox{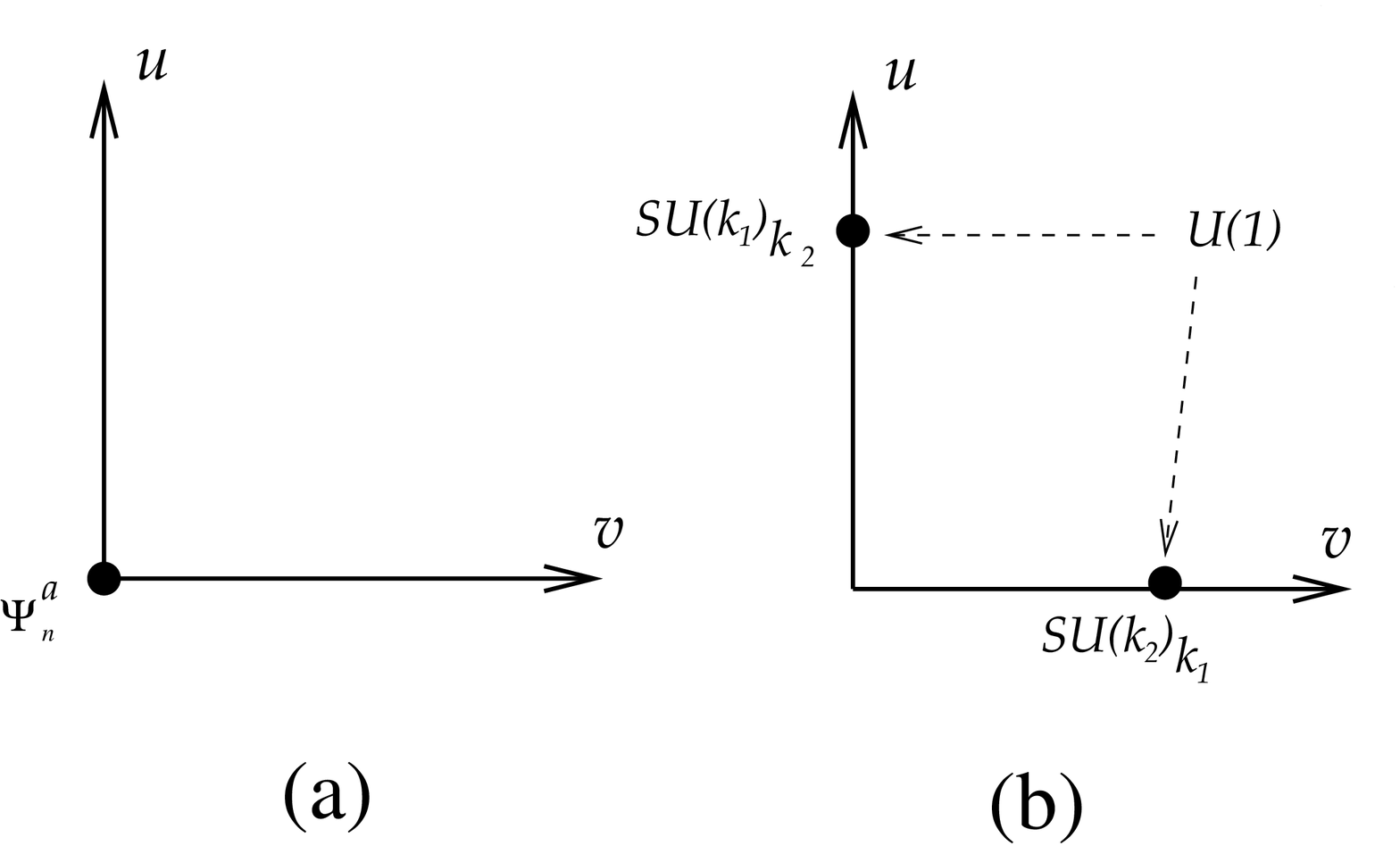}}

As the coupling constant is increased, the current algebra moves
to infinity and the low energy theory around the real axis in
figure 1 becomes very simple.  First, the system has a gap there.
Second, the topological degrees of freedom associated with the
Chern-Simons theories in the lower and upper half planes combine
nicely.  We have $SU(k_1)$ Chern-Simons theory with level $k_2$ in
the upper half plane and $SU(k_2)$ level $k_1$ in the lower half
plane. Due to level-rank duality, these two theories are the same.
Therefore, the low energy dynamics of this brane configuration is
a Chern-Simons theory with gauge group $SU(k_1)_{k_2}\times U(1)$
both in the lower and in the upper half plane. The ``domain wall''
along the real axis disappears and our low energy theory is fully
invariant under $2+1$ dimensional Poincare symmetry.

The existence of this $2+1$ dimensional Poincare symmetry has
immediate implications to the supersymmetry of the system. The
eight supercharges \conssusy\ do not fit into a representation of
this symmetry. As the Poincare symmetry is enhanced from
two dimensions to three, the supersymmetry is also enhanced from
eight to sixteen supercharges. The superalgebra is not the
standard one -- the anticommutator of two supercharges includes in
addition to the momentum operator also the generators of the
$SO(4)\times SO(4)$ R-symmetry of the problem.\foot{It is not
known how to write a three dimensional Lagrangian for gauge fields
with Chern-Simons terms and this much supersymmetry (for a related
recent discussion see \SchwarzYJ ). In the past, the search for
such Lagrangians was based on the standard extended supersymmetry
algebra. The string theory problem we are studying suggests that
with the unconventional superalgebra that we find, Lagrangians
with sixteen supercharges and Chern-Simons terms can be written.
It would be interesting to see whether kinetic terms for the gauge
fields can be included in such Lagrangians. This is not entirely
clear from the string theory perspective, since at the scale of
the gauge coupling our theory becomes higher dimensional, but we
believe that it is nevertheless possible.}

\subsec{Strong coupling analysis}

In section 4 we turn to the strong coupling limit of our system.
Here it is convenient to perform an S-duality transformation to a
system of $NS5$-branes.\foot{The discussion can be extended to any
other kind of $(p,q)$ fivebranes. The $SL(2,Z)$ symmetry of IIB
string theory implies that all such configurations have the same
amount of supersymmetry, low lying spectrum and dynamics. In
particular, the chiral fermions, the gauge fields and the action
that couples the two are the same for all $(p,q)$. The only part
of the gauge theory analysis that depends on $(p,q)$ is the value
of the gauge coupling on the fivebranes. Since the gauge coupling
determines the extent by which the holomorphic currents \curralg\
are displaced from the intersection of the branes, the latter
depends on the type of fivebranes that are used in the
construction.} When all the fivebranes are coincident, the
near-horizon geometry is \eqn\mh{\IR^{2,1}\times \IR_\phi\times
SU(2)_{k_1}\times SU(2)_{k_2}. } Here, $\IR_\phi$ is one
combination of the radial directions away from the two sets of
fivebranes, and the coordinates of $\IR^{2,1}$ are $x^0$, $x^1$
and  another combination of the two radial directions. The two
$SU(2)$'s describe the angular three-spheres corresponding to
$\left(\IR^4\right)_{2345}$ and $\left(\IR^4\right)_{6789}$. The
fact that \mh\ is an exact solution of the classical string theory
equations of motion allows us to obtain information about the
intersecting fivebrane system, which is not accessible via a gauge
theory analysis.

As mentioned above, the geometry \mh\ has the interesting property
that it exhibits a higher symmetry than the full brane
configuration.  In particular, the combination of radial
directions away from the intersection that enters $\IR^{2,1}$
appears symmetrically with the other spatial direction, and the
background has a higher Poincare symmetry, $ISO(2,1)$, than the
expected $ISO(1,1)$. It also has twice as much supersymmetry.

A natural question is whether this higher symmetry is  an exact
property of  string theory in the background \mh, or whether it is
broken by quantum effects. In weakly coupled string theory,
classical symmetries are usually symmetries of the full theory,
but the theory on \mh\ is not weakly coupled. As $\phi\to-\infty$,
the string coupling becomes strong and one could  imagine that in
that region the symmetry of \mh\ is broken to a smaller one.

A similar divergence appears in the six dimensional background
corresponding to a single stack of $k$ parallel fivebranes in type
II string theory. In that case, the region $\phi\to-\infty$
describes the long distance behavior of the theory on the branes
\ItzhakiDD,
which has an alternative description as a six dimensional $U(k)$
SYM theory with sixteen supercharges for type IIB string theory,
and as the $(2,0)$ SCFT for IIA. In both cases, the low
energy theory has non-Abelian degrees of freedom which from the
point of view of the geometry are D-branes living near the
singularity. The fact that they are light implies that string
perturbation theory must break down there.

It is natural to follow the same logic in our case. The background
\mh\ becomes strongly coupled near the intersection of the branes
partly due to the fact that the low energy dynamics near the
intersection involves a non-Abelian gauge theory. Thus, to see
whether the $2+1$ dimensional super-Poincare symmetry is broken in
the strong coupling region, one needs to analyze the gauge theory.

The results of this analysis are described in the previous subsection
(and in more detail in sections 2, 3). The gauge theory dynamics displaces
the chiral fermions, which naively live at the intersection, by an
amount of order $g_{YM}$, which for $NS5$-branes is $l_s$. Since
the near-horizon limit that gives rise to \mh\ focuses on
distances much smaller than $l_s$, the chiral fermions are not
part of the theory in the near-horizon limit. In fact, in taking
the near-horizon limit, each of the three factors in \curralg\ is
sent to infinity in a different direction. The two $SU(k_i)$ are
sent to plus and minus infinity in one of the two spatial
directions of $\IR^{2,1}$, while the $U(1)$ is sent to infinity in
the weak coupling direction of $\IR_\phi$.

So far we restricted attention to the extreme infrared behavior of
the intersecting fivebrane system, and found that it is consistent
with Poincare symmetry in $2+1$ dimensions. It is natural to
postulate that a much stronger statement, that we did not prove,
is true: the near-horizon limit eliminates {\it all} $1+1$
dimensional degrees of freedom, and leaves behind a $2+1$
dimensional theory with a larger symmetry. We present some
evidence for this  in section 6.

\subsec{The big picture}

The discussion above focused on the dynamics associated with the
intersection in the case when the fivebranes are coincident. Some
new features appear when we separate the two sets of fivebranes in
the corresponding transverse $\IR ^4$'s. One consequence of this
separation is a breaking of the Poincare symmetry of the
near-horizon geometry back to $ISO(1,1)$, and of half of the
sixteen supercharges, back to the original eight.  The
$U(k_i)$ gauge groups are generically broken to $U(1)^{k_i}$ and
an additional dimensionless parameter, the mass of the W-bosons
of the broken gauge theory in units of the inverse gauge coupling
($M_s$ for $NS5$-branes), appears.

For $M_W\ll M_s$, the deformed near-horizon geometry which
generalizes \mh\ is still strongly coupled, and the useful
description of the system is given by the gauge theory described
earlier. On the other hand, for $M_W\gg M_s$, the gauge theory
description is in general not valid\foot{Although, as in
\AharonyXN, it may still be useful for some purposes.}, and the
useful description is in terms of the near-horizon geometry, a
deformation of \mh, which never develops large string
coupling. By analyzing the bulk geometry we show
that all $k_1 k_2$ chiral modes are present, as certain chiral
modes of RR fields in the geometry, and find their locations in
the transverse space.

In section 5 we combine the information obtained from the gauge
theory analysis for $M_W\ll M_s$, with that obtained from the
near-horizon geometry of the fivebranes for $M_W\gg M_s$ to
reconstruct the behavior of  the chiral modes as we vary the
different parameters. In section 6 we point out a few extensions
of our work and directions for future study.
Some of our worldsheet computations are presented in the
appendix.

\newsec{Gauge theory analysis for $k_1=k_2=1$}

In this section we will study the low energy field theory
corresponding to the intersection of two $D5$-branes, one
stretched in the directions $(012345)$, the other in $(016789)$.
As discussed in the introduction, this system contains a single
complex chiral fermion, $\Psi$, that lives at the intersection.
This fermion carries equal and opposite charges under the $U(1)$
gauge fields living on the  two intersecting fivebranes. We will
describe the combined system of the fermion and the gauge fields,
and in particular analyze the effects of anomaly inflow on the
dynamics.

 \subsec{The Lagrangian}

It is useful to parameterize the ten dimensional spacetime as
follows:
\eqn\met{ds^2=2dx^+ dx^- + du^2 + u^2d\Omega_u^2+dv^2 + v^2 d\Omega_v^2.}
Here, $(u,\Omega_u)$ are spherical coordinates on
$\left(\IR^4\right)_{2345}$ and $(v,\Omega_v)$ are spherical
coordinates on $\left(\IR^4\right)_{6789}$. The two intersecting
fivebranes share the directions $x^\pm=(x^0\pm x^1)/\sqrt2$; in
addition, one is wrapped on $(u,\Omega_u)$, while the other is
wrapped on $(v,\Omega_v)$. The two fivebranes intersect at
$u=v=0$.

The low energy gauge theory includes a $U(1)$ gauge field that
lives on the worldvolume of each kind of fivebrane, which we will
refer to as $A^{(1)}$ and $A^{(2)}$, respectively. $A^{(1)}$ lives
in the six dimensional spacetime labeled by $(x^\pm, u,
\Omega_u)$, while  $A^{(2)}$ is a function of $(x^\pm, v,
\Omega_v)$. We can expand each of the gauge fields in harmonics on
the corresponding three-sphere. Since we are only interested in
the lowest lying states living near the intersection, we integrate
out the higher harmonics and keep in the effective action only the
s-waves of $A^{(i)}$ on the spheres.

We expect the low energy effective action near the intersection to
be a sum of three terms: a $2+1$ dimensional action for the field
$A^{(1)}(x^\pm,u)$, the s-wave of $A^{(1)}$ on the sphere labeled
by $\Omega_u$; a similar $2+1$ dimensional action for the s-wave
of $A^{(2)}$, $A^{(2)}(x^\pm,v)$; and a $1+1$ dimensional action
which lives at the intersection $u=v=0$. In the rest of this
subsection we will write this action explicitly.

We start with the $1+1$ dimensional action associated with the
intersection. The complex fermion $\Psi$  couples to the gauge
field $A^{(1)}-A^{(2)}$ at the intersection point, $u=v=0$, which
we will denote by $A^{(1)}(0)-A^{(2)}(0)$, suppressing the
dependence on $x^\pm$. As is standard in two dimensional field
theory, one can integrate out the fermion and represent its
dynamics in terms of its contribution to the effective action of
the gauge field,\foot{The overall coefficient of the two
dimensional Lagrangian implies a choice of normalization of the
charge of the fermion under the $U(1)$ gauge fields $A^{(i)}$, or
equivalently a choice of normalization of the gauge fields
themselves. This determines the coefficients of all the other
terms that appear below.}
 \eqn\lferm{\eqalign{ {\cal L}_{\rm ferm}=&
 \left(A_+^{(1)}(0)-A_+^{(2)}(0)\right) {\partial_-\over\partial_+}
 \left(A_+^{(1)}(0)-A_+^{(2)}(0)\right) \cr-&
 \left(A_+^{(1)}(0)-A_+^{(2)}(0)\right)
 \left(A_-^{(1)}(0)-A_-^{(2)}(0)\right). }}
This effective Lagrangian is anomalous. Under the gauge transformation
 \eqn\gauget{\delta
 A_\mu^{(i)}=\partial_\mu\epsilon^{(i)},~~~~~i=1,2,}
${\cal L}_{\rm ferm}$ transforms as follows
 \eqn\varlferm{\delta {\cal L}_{\rm ferm}=
 \left(\epsilon^{(1)}(0)-\epsilon^{(2)}(0)\right)
 \left(F_{+-}^{(1)}(0)-F_{+-}^{(2)}(0)\right). }
As explained in \ghm,  this anomaly is cancelled by inflow from
the bulk of the two fivebranes, via local terms in the Lagrangians
on the two branes.

There are two relevant types of terms. One involves the coupling
of the Chern-Simons form of the gauge field $A^{(i)}$ to the RR
three-form field strength, $F_3$. For example, on brane $1$, this
takes the form\foot{We are not careful with the overall
normalization here; we will determine it in our conventions
momentarily, after reducing to $2+1$ dimensions.}
\eqn\qp{ \int F_3 \wedge A^{(1)}\wedge F^{(1)}. }
The three-sphere labeled by $\Omega_u$ on which fivebrane $1$ is
wrapped  is transverse to fivebrane $2$. Therefore, there is one
unit of $F_3$ flux going through it. Thus after integrating over
the sphere, we find a three dimensional Chern-Simons action for
the s-wave of $A^{(1)}$, $A^{(1)}(x^\pm,u)$, and a similar one for
$A^{(2)}(x^\pm,v)$:
\eqn\csaction{\eqalign{{\cal L}_{\rm CS}=&\int_0^\infty du\left(
A^{(1)}_+F^{(1)}_{-u}+ A^{(1)}_-F^{(1)}_{u+}+
A^{(1)}_uF^{(1)}_{+-} \right)\cr +&\int_0^\infty dv\left(
A^{(2)}_+F^{(2)}_{-v}+ A^{(2)}_-F^{(2)}_{v+}+
A^{(2)}_vF^{(2)}_{+-} \right) .}}
As is standard in Chern-Simons theory with a boundary, the Lagrangian
\csaction\ is not gauge invariant. Under the gauge transformation \gauget\
it transforms as follows:
\eqn\cstra{\delta {\cal L}_{\rm CS}
=-\epsilon^{(1)}(0)F_{+-}^{(1)}(0)-\epsilon^{(2)}(0)F_{+-}^{(2)}(0),}
where we used the fact that we only require invariance under transformations
that go to zero at infinity, \ie\  the gauge parameters satisfy
$\epsilon^{(i)}(\infty)=0$.

The second type of Chern-Simons term that is relevant for anomaly
inflow involves a coupling of the worldvolume gauge fields to the
self-dual RR field strength, $F_5$,
\eqn\pq{ \int F_5 \wedge A^{(i)}.}
Since the D5-branes do not couple to $F_5$, naively this term is
irrelevant for our discussion. However when the gauge fields on
both branes are turned on, this term has a non-trivial effect. For
example, if we turn on a non-zero gauge field on brane 2, with field
strength $F^{(2)}_{+-}$, this induces a non-zero $F_5$ on brane 1 of
the form
\eqn\yu{F_5=F_3\wedge  F^{(2)}(0)~.}
Here, $F_3$ is the RR three-form field strength induced by brane 2
on the sphere $\Omega_u$, as in \qp, and $ F^{(2)}$ is evaluated
at $v=0$, since this is the location of brane 1. Substituting \yu\
into \pq, reducing to s-waves, and integrating over the spheres,
one finds the following contribution to the action:
\eqn\mixac{{\cal L}_{\rm mix} =-F^{(2)}_{+-}(0)\int_0^\infty du
A^{(1)}_u -F^{(1)}_{+-}(0)\int_0^\infty dv A^{(2)}_v .}
The total anomaly inflow Lagrangian is given by
\eqn\linflow{ {\cal L}_{\rm inflow}={\cal L}_{\rm CS}+{\cal
L}_{\rm mix}~.}
The full action,
\eqn\actionuone{S_{\rm 0}=\int dx^+ dx^- \left( {\cal L}_{\rm
ferm}+{\cal L}_{\rm inflow}\right) ,}
is anomaly free.\foot{Here we restrict to gauge anomalies,
and ignore gravitational ones.}

In addition to the fermion contribution and anomaly inflow Lagrangian,
we also need to keep the kinetic terms of $A^{(1)}$ and $A^{(2)}$. After
reducing on the spheres, these take the form
\eqn\lkin{\eqalign{ {\cal L}_{\rm kin}= &{1\over
g_1^2}\int_0^\infty du u^3
\left[{1\over2}\left(F_{+-}^{(1)}\right)^2-F^{(1)}_{u+}F^{(1)}_{u-}\right]\cr
+&{1\over g_2^2}\int_0^\infty dv v^3
\left[{1\over2}\left(F_{+-}^{(2)}\right)^2-F^{(2)}_{v+}F^{(2)}_{v-}\right]
.\cr  }}
Here $g_1,g_2$ are the six dimensional gauge couplings, $g_i^2\simeq g_sl_s^2$
for $D5$-branes, or $g_i^2\simeq l_s^2$ for $NS5$-branes, and we have absorbed
a numerical factor in the definition of $g_i$. In principle, one should also add higher
derivative terms to \actionuone, but we expect them not to alter the picture obtained below.

\subsec{Equations of motion and solutions}

The full effective action
\eqn\act{S_{\rm full}=\int dx^+dx^- \left({\cal L}_{\rm ferm}+
{\cal L}_{\rm inflow}+{\cal L}_{\rm kin} \right)}
is quadratic in the gauge fields; therefore, to solve the theory
it is enough to study solutions to the equations of motion of \act.
In this subsection we will write down these equations and find
the solution corresponding to the chiral mode associated with
the fermion $\Psi$.

Due to the presence of the boundary at $u=v=0$, we have to
vary separately with respect to the gauge field in the bulk
and on the boundary. Varying with respect to $A_+^{(1)}$ in
the bulk (\ie\ for generic $u$) we find
\eqn\aazz{{1\over g_1^2}\partial_u\left(u^3F^{(1)}_{u-}\right)
-2F^{(1)}_{u-}+{u^3\over g_1^2}\partial_-F_{+-}^{(1)}=0.}
Varying with respect to $A^{(1)}_-$ gives
\eqn\aabbzz{{1\over g_1^2}\partial_u\left(u^3F^{(1)}_{u+}\right)
+2F^{(1)}_{u+}+{u^3\over g_1^2}\partial_+F_{-+}^{(1)}=0,}
and the variation of $A_u^{(1)}$ gives
\eqn\aavv{2F^{(1)}_{+-}-F^{(2)}_{+-}(0)- {u^3\over
g_1^2}\left(\partial_-F^{(1)}_{u+}+
\partial_+F^{(1)}_{u-}\right)=0.}
Similar equations are obtained with $1\leftrightarrow 2$,
$u\leftrightarrow v$.

Due to the presence of the Chern-Simons term we expect the
equations of motion \aazz\ -- \aavv\ to describe a massive
particle in three dimensions with a mass that depends on $u$
(because of the factor of $u^3$ in ${\cal L}_{\rm kin}$). Indeed,
differentiating equation \aazz\ w.r.t. $x^+$, and equation
\aabbzz\ w.r.t. $x^-$, and subtracting the two we find, using
\aavv\
\eqn\massive{
\partial_u\left(u^3\partial_u F_{+-}^{(1)}\right)+
2u^3\partial_+\partial_-F_{+-}^{(1)}- {2g_1^4\over
u^3}\left(2F_{+-}^{(1)}-F_{+-}^{(2)}(0)\right)=0.}
The mass of the particle goes to zero as $u\to\infty$, but it does
not give rise to massless degrees of freedom in the $\IR^{1,1}$
labeled by $x^\pm$. One can show that equation \massive\ does not
have solutions with $\partial_+\partial_-F_{+-}^{(1)}=0$ that are
regular as $u\to 0$ and go to zero as $u\to\infty$. The massless
regular solutions of \massive\ go to a non-zero constant at
$u=\infty$ and correspond to a constant electric field in
$\IR^{1,1}$. In other words, \massive\ describes a massive degree
of freedom  in the bulk of fivebrane $1$.

Thus, we set
\eqn\setfpm{F^{(1)}_{+-}=F^{(2)}_{+-}=0.}
Equations \aazz\ and \aabbzz\ can then be solved for
$F_{u\pm}^{(1)}$
\eqn\solfpm{F_{u\pm}^{(1)}={h_\pm\over u^3} e^{\pm {g_1^2\over u^2
}} ,}
where $h_{\pm}$ depend on $x^\pm $ but not on $u$. While the
solution for $F_{u-}^{(1)}$ is well behaved as $u\to 0,\infty$,
the solution for $F_{u+}^{(1)}$ is highly divergent as $u\to 0$,
so we discard it, and set $h_+=0$. Moreover, substituting in
\aavv\ we see that $h$ satisfies the constraint
\eqn\chirh{\partial_+ h_-=0~.}
Thus, it is a chiral degree of freedom, $h_-=h_-(x^-)$.

Next we consider the equations of motion of the boundary
variables. Varying w.r.t. $A^{(1)}_+(0)$ we get
\eqn\boundaazz{ \int_0^\infty dv F_{-v}^{(2)}
={2\over\partial_+}\left(F_{-+}^{(1)}(0)
-F_{-+}^{(2)}(0)\right)\equiv g(x^-).}
In deriving this equation we used the fact that $\lim_{u\to 0}u^3
F_{u-}^{(1)}=0$, \solfpm. Note also that the quantity $g(x^-)$
defined by \boundaazz\ is indeed chiral, $\partial_+ g=0$, due
to \setfpm. Again, there are two more equations obtained by
exchanging $1\leftrightarrow2$.

To recapitulate, we see that including the interaction of the
chiral fermion $\Psi$ with the gauge fields living on the two
fivebranes via anomaly inflow displaces it from the intersection
of the two branes at $u=v=0$ by an amount of order the gauge
coupling on the fivebranes. One finds a single chiral degree of
freedom, which can be thought of as $h_-(x^-)$ \solfpm, that lives
at $u\sim g_1$. Equation \boundaazz\  relates it to the current
located at the origin. The equations of motion on the second brane
relate this mode to a solution similar to \solfpm\ that lives at
$v\sim g_2$. The situation is depicted in figure 2(b).

At short distances from the intersection, $u,v\ll g_i$, the
fermions are absent and the physics is described by $U(1)$
Chern-Simons theory on $\IR^{2,1}$.

\newsec{Gauge theory analysis for general $k_1$, $k_2$}

A natural generalization of the brane configuration discussed in
the previous section involves $k_1$ fivebranes stretched in
$(012345)$, intersecting $k_2$ fivebranes stretched in $(016789)$.
At the intersection we have $k_1k_2$ complex fermions $\Psi_n^a$;
$n=1,\cdots, k_1$; $a=1,\cdots, k_2$. The fermions are coupled to
$U(k_1)\times U(k_2)$ gauge fields. As mentioned in the
introduction, one can construct out of the fermions left-moving
currents of the affine Lie algebra $SU(k_1)_{k_2}\times
SU(k_2)_{k_1}\times U(1)$. As a check, the Sugawara central charge
of the current algebra agrees with that of the fermions:
\eqn\centagree{ { k_2 (k_1^2-1) \over k_1+k_2}+ { k_1 (k_2^2-1)
\over k_1+k_2}+1=k_1 k_2 ~.} In generalizing the discussion of the
previous section to this case there are two new issues to
consider. One is that the gauge theory in question is now
non-Abelian, so the dynamics is more complicated. The other is
that the brane configuration admits deformations that correspond
to separating each stack of fivebranes in the transverse $\IR^4$,
and one may ask how the low energy behavior of the theory depends
on the separations.

In this section we will examine these issues in turn. We start with a discussion of
the dynamics for the case of coincident fivebranes, and then move on to the
separated case.

\subsec{Coincident fivebranes}

Like in the Abelian case, we integrate out the fermions, and replace
them by their contribution to the effective action of the gauge fields.
This action takes the  form
 \eqn\lfnonab{
{\cal L}_{\rm ferm}=k_2\Gamma(A^{(1)}(0))+k_1\Gamma(A^{(2)}(0))
+\Gamma(U(1)),}
where the $U(1)$ is the relative $U(1)$ in $U(k_1)\times U(k_2)$,
and $\Gamma(A)$ is the Polyakov-Wiegmann action \pw\ corresponding
to the relevant gauge group. The overall $U(1)$ does not appear in
${\cal L}_{\rm ferm}$ since it does not couple to the  fermions.
$\Gamma(U(1))$ is the action \lferm\ with traces taken over the
two gauge groups. The $U(1)\times U(1)$ part of the dynamics
decouples from the non-Abelian parts. It is identical to the
discussion in the previous section. So we  ignore it below, and
focus on the $SU(k_1)\times SU(k_2)$ part.

We need the non-Abelian generalization of ${\cal L}_{\rm kin}$
\lkin\ and ${\cal L}_{\rm inflow}$ \linflow. The kinetic terms for
the gauge fields are simple generalizations of \lkin. The
generalization of \qp\ is
\eqn\qpna{ \int F_3 \wedge  CS(A^{(i)})~,}
where $CS(A)$ is the Chern-Simons form,
\eqn\csdef{CS(A^{(i)})\equiv\Tr \left(A^{(i)} \wedge F^{(i)} +\frac23
A^{(i)}\wedge A^{(i)} \wedge A^{(i)}\right)~. }
There are $k_1$ units of $F_3$ flux through the sphere labeled by
$\Omega_v$ \met,  and $k_2$ units of $F_3$ through the sphere
$\Omega_u$.  Integrating over the three-spheres, as in the
previous section, we find that the non-Abelian part of ${\cal
L}_{\rm CS}$ reads
\eqn\linnonab{{\cal L}_{\rm CS}=k_2CS(A^{(1)})+k_1CS(A^{(2)})~.}
The generalization of \pq\ is
\eqn\pqna{ \int F_5 \wedge \Tr A^{(i)}~.}
Therefore, $F_5$ only couples to the $U(1)$ parts of the gauge
groups, and can be ignored in the present discussion. Note that
the Chern-Simons terms do not mix the non-Abelian gauge fields on
the two branes. The $SU(k_1)\times SU(k_2)$ anomalies cancel
between the fermion contribution \lfnonab\ and the Chern-Simons
term  \linnonab.

The low energy dynamics breaks up into two decoupled problems, one
for $SU(k_1)$ and the other for $SU(k_2)$. Thus, it is enough to
discuss one of them, say that of $SU(k_1)$. The effective two
dimensional Lagrangian for the gauge field $A^{(1)}_\mu$ is
\eqn\llun{{\cal L}_1=k_2\Gamma(A^{(1)}(0))+k_2\int_0^\infty du\
CS(A^{(1)})+ {1\over g_1^2}{\rm Tr}\ \int_0^\infty duu^3
\left[{1\over2}\left(F^{(1)}_{+-}\right)^2-F^{(1)}_{u+}F^{(1)}_{u-}\right]
.}
 The
Lagrangian is no longer quadratic in the fields, but it is still
instructive to look at the solutions of the equations of motion.
Again we are interested in massless excitations in two dimensions,
so we set (as in \setfpm\ and the discussion following it)
\eqn\sepnew{ F^{(1)}_{+-}(u)=0,~~~~ D_-F^{(1)}_{u+}=D_{+}F^{(1)}_{u-}=0,}
where $D$ stands for covariant derivatives in $SU(k_1)$. The
analog of the bulk equations of motion, \aazz, \aabbzz\ is
\eqn\aanew{\eqalign{
{1\over g_1^2}D_u\left(u^3F^{(1)}_{u-}\right)-2k_2 F^{(1)}_{u-}=&0,\cr
{1\over g_1^2}D_u\left(u^3F^{(1)}_{u+}\right)+2k_2 F^{(1)}_{u+}=&0~. }}
To simplify these equations it is convenient to choose the gauge
$A^{(1)}_u=0$, such that $D_u=\partial_u$ and $F^{(1)}_{u\pm}=\partial_u
A^{(1)}_\pm$. As in section 2, we conclude that $F^{(1)}_{u+}=0$ and
\eqn\formaz{A^{(1)}_-=h_-(x^-)e^{-{g_1^2 k_2 \over u^2}}~.}
We see that the $SU(k_1)$ dynamics describes a chiral $k_1\times
k_1$ matrix $h_-(x^-)$, which lives at $u\sim g_1 \sqrt{k_2}$.
In contrast to section 2, this matrix does not consist of free
fields. The non-Abelian dynamics of \llun\ gives rise to
interactions, such that $h_-$ in fact describes a chiral $SU(k_1)$
WZW model at level $k_2$. Similarly, the $SU(k_2)$ dynamics gives
rise to a chiral $k_2\times k_2$ matrix living at
$v\sim g_2\sqrt{k_1}$, describing a chiral WZW model
$SU(k_2)_{k_1}$ (see figure 2(b)).

The fact that the $SU(k_1)$ and $SU(k_2)$ currents commute, and
can be completely decoupled can be seen in the gauge theory as
follows. Varying the Lagrangian \llun\ with respect to the
boundary variables, we get an analog of equation \boundaazz, but
with zero on the left hand side. The reason is that the origin of
the left hand side in \boundaazz\ is ${\cal L}_{\rm mix}$ \mixac,
which is absent in the non-Abelian case. Thus, the non-Abelian
analog of the function $g(x^-)$, defined in \boundaazz, vanishes
and the two non-Abelian parts of the current algebra do not couple
to each other.

Note that the fact that the $SU(k_1)$ and $SU(k_2)$ currents
commute and live at different locations in the plane (figure 2(b))
does not mean that the two chiral WZW models, $SU(k_1)_{k_2}$ and
$SU(k_2)_{k_1}$ are completely decoupled. In fact, each of them
separately is not a consistent quantum field theory. Many physical
observables, such as the fermions $\Psi_n^a$, must have
support in all parts of the $(u,v)$ plane in which the currents
are localized, since only when we combine the contributions from
the two $SU(k_i)$ and from $U(1)$ do we get single valued
observables in two dimensions.

To summarize, we arrive at the following picture. At $g_s=0$, the
$k_1 k_2$ chiral fermions $\Psi_n^a$ are located at the
intersection, $u=v=0$, as suggested by their description in terms
of open strings stretched between $D5$-branes. Turning on the
coupling constant does not lift them (which is impossible due to
their chirality), but rather moves them away from the origin.
Different currents constructed out of the fermions move in
different directions (in the way depicted in figure 2). The
$SU(k_1)_{k_2}$ currents move to $u\sim g_1\sqrt{k_2}$. The
$SU(k_2)_{k_1}$ ones move to $v\sim g_2\sqrt{k_1}$. The $U(1)$
part is supported in both regions.

The fermions themselves, which can be thought of as solitons
constructed out of the currents, become delocalized, ``fat,''
objects, with support in both regions mentioned in the previous
paragraph. Thus, the main physical effect of the anomaly inflow is
to delocalize the two dimensional physics in the higher
dimensional space.

Focusing on the region near the intersection, $u\ll g_1$, $v\ll g_2$,
we find no massless degrees of freedom. In this region, the extreme
IR theory on the first type of fivebranes is $SU(k_1)$ Chern-Simons
theory at level $k_2$, living on the half plane $u\geq 0$, while that
on branes of the second type has gauge group $SU(k_2)$ and level
$k_1$, and lives on the half plane $v\geq 0$. The two boundaries of
the half-planes, $u=v=0$ are identified (see figure 1).

Level-rank duality implies that the two Chern-Simons theories living on
the $u$ and $v$ half-planes are the same, so in fact the infrared theory
can be described as a {\it single} Chern-Simons theory (either
$SU(k_1)_{k_2}$ or $SU(k_2)_{k_1}$) living on the whole plane, or,
after including time, on $\IR ^{2,1}$.

\subsec{Separated fivebranes}

In this subsection we discuss the modification of the gauge theory
picture when  the fivebranes are separated in the transverse
$\IR^4$'s. The main effect of the separation is to give masses to
some of the gauge bosons, generically breaking the gauge symmetry
from $U(k_i)\to U(1)^{k_i}$. For concreteness, and to facilitate
the detailed comparison to the closed string analysis of the next
section, we will focus on a point in moduli space, at which each
set of fivebranes is arranged symmetrically around a circle of
radius $r$ in the transverse $\IR^4$. It is not difficult to
generalize the analysis to more general fivebrane configurations;
the qualitative picture is independent of the precise pattern of
separations.

In the $NS5$-brane configuration we will discuss, the masses of all the
W-bosons are comparable, of order $M_W\simeq rM_s^2/g_s$. We will
assume that $M_W\ll M_s$. This corresponds to very
small separations of the fivebranes, $r\ll g_sl_s$. The regime
$M_W\gg M_s$ will be analyzed in the next section, by studying the
geometry created by the fivebranes.

How does a small $M_W$ influence the picture arrived at in the
previous subsection? We expect that different parts of the current
algebra will react differently to the deformation. Since a
$U(1)^{k_1-1}$ out of the $SU(k_1)$ that lives at
$u\sim g_1\sqrt{k_2}$ is unbroken, we expect
it to remain where it was, while the rest of the currents,
parameterizing the coset $SU(k_1)_{k_2}/U(1)^{k_1-1}$ will presumably
react in some way. The same should be true for $SU(k_2)_{k_1}$.

To analyze this problem precisely, one must solve the full
non-Abelian gauge theory for finite $M_W$. This is difficult,
among other things because after separating the fivebranes one can
no longer reduce the problem to three dimensions, as we have done
in the previous subsection. Rather, one has to study the full six
dimensional theory on each set of fivebranes.

Here we will content ourselves by studying the theory at large
distances, at which the massive gauge bosons are irrelevant, and
one can restrict to the dynamics of the Abelian gauge theory,
$U(1)^{k_1+k_2}$. This description is expected to be valid for
$u,v\gg 1/M_W$. Thus, it can  provide information only about the
long distance behavior of the various modes.

At distances much larger than $1/M_W$, the analysis of the previous
subsection goes through, the only difference being that we take the
gauge fields $A^{(i)}$ to be diagonal matrices. For example,
$h_-(x^-)$ in \formaz\ is a diagonal $k_1\times k_1$ matrix, with
eigenvalues $h_n(x^-)$, $n=1,2,\cdots k_1$. Similarly, the corresponding
$k_2\times k_2$ matrix for $A^{(2)}$ is diagonal with eigenvalues
$\tilde h_a(x^-)$, $a=1,\cdots, k_2$.

Thus, in this case we have $k_1k_2$ copies of the analysis of
section 2. In particular, for every pair $(n,a)$ we find a single
chiral degree of freedom, which can be interpreted as $h_n(x^-)$
or as $\tilde h_a(x^-)$. The two are identified for the same reason
as in section 2.  From the point of view of the original fermions
$\Psi_n^a$, these degrees of freedom are the $U(1)$ currents
\eqn\jjiiaa{J_n^a=\left(\Psi^*\right)^n_a\Psi_n^a~.}
They correspond to wavefunctions that decay like $1/u^3$ and $1/v^3$
at large distances from the intersection, as in section 2.

As mentioned in the beginning of this subsection, we expect the
currents \jjiiaa\ to split naturally into a number of groups,
\eqn\splitgroups{U(1)\times U(1)^{k_1-1}\times U(1)^{k_2-1}
\times \left[{SU(k_1)_{k_2}\over U(1)^{k_1-1}}\times
{SU(k_2)_{k_1}\over U(1)^{k_2-1}}\right]~.}
The first three factors in \splitgroups\ correspond to the unbroken
parts of the gauge symmetry, while the factor in square
brackets is the part of the theory that is expected to react
to the separation of the branes.

In terms of the currents  $J_n^a$, \jjiiaa, the different factors
in \splitgroups\ can be written as follows. The $U(1)^{k_1+k_2-1}$
currents can be written as
 \eqn\uonecur{\eqalign{ U(1):\;\;\;
 &\sum_{n,a}J_n^a~,\cr U(1)^{k_1-1}:\;\;\; &\sum_{n,a}J_n^a e^{2\pi
 in(2j_1+1)\over k_1}~,\cr U(1)^{k_2-1}:\;\;\; &\sum_{n,a}J_n^a
 e^{2\pi ia(2j_2+1)\over k_2}~,\cr }}
where
 \eqn\rangejj{2j_i=0,1,2,\cdots, k_i-2~.}
The remaining currents, corresponding to the cosets in the square
brackets in \splitgroups, are the combinations of the $J_n^a$ that
are orthogonal to all those in \uonecur. A convenient basis for
them is \eqn\basrest{\tilde J_{j_1, j_2}=\sum_{n,a} J_n^a e^{2\pi
in(2j_1+1)\over k_1} e^{2\pi ia(2j_2+1)\over k_2}~,} where $j_i$
run over the range \rangejj.

The splitting of the $k_1k_2$ currents $J_n^a$ into the groups
\splitgroups\ is natural from the point of view of our gauge theory
analysis. Different groups correspond to different asymptotic decay
of the field strengths $F^{(1)}$ at large $u$ and $F^{(2)}$ at large $v$.
The overall $U(1)$ degree of freedom corresponds to
\eqn\decayuone{F^{(1)}(u)\sim {1\over u^3};\;\;\; F^{(2)}(v)\sim {1\over v^3}~,}
the $U(1)^{k_1-1}$ currents \uonecur\ correspond to
\eqn\uonekone{F^{(1)}(u)\sim {1\over u^3};\;\;\; F^{(2)}(v)\sim {1\over v^{2j_1+4}}~,}
the $U(1)^{k_2-1}$ currents correspond to
\eqn\uonektwo{F^{(1)}(u)\sim {1\over u^{2j_2+4}};\;\;\; F^{(2)}(v)\sim {1\over v^3}~,}
and the remaining currents \basrest\ have a faster than inverse
cubed decay in both $u$ and $v$
\eqn\cosss{F^{(1)}(u)\sim {1\over u^{2j_2+4}};\;\;\; F^{(2)}(v)\sim {1\over v^{2j_1+4}}~.}
It should be noted that the precise form of the asymptotic fall-off
\decayuone\ -- \cosss\ is valid for the particular point in the moduli space
where the fivebranes are placed on circles in the transverse $\IR^4$'s,
but the basic fact that the $U(1)^{k_1+k_2-1}$ currents correspond to
field strengths that decay like $1/u^3$ and/or $1/v^3$, while the factor
in square brackets in \splitgroups\ corresponds to field strengths with
more rapid decay in both $u$ and $v$, is general.

It is also worth reiterating that the Abelian analysis above only
captures the behavior of the different modes for large $u$, $v$.
It is not sensitive enough to see how figure 2 gets deformed as we
turn on a small $M_W$; that requires tools with much higher
resolution. In particular, the fact that the factor in square
brackets in \splitgroups\ consists (for $NS5$-branes) of a part
that lives near $v=l_s\sqrt{k_1}$, and a part that lives near
$u=l_s\sqrt{k_2}$, as is expected from the analysis that led to
figure 2, cannot be resolved at this level. The Abelian analysis
combines these two parts together and replaces them by the
$(k_1-1)(k_2-1)$ currents \basrest.

\newsec{Closed string description}

The supergravity background corresponding to the configuration of
intersecting $NS5$-branes discussed in the previous sections
factorizes into three decoupled parts \khu. The $\IR^{1,1}$,
labeled by $x^\pm$, which is common to both sets of branes, is
trivial. The four dimensional space labeled by $(x^2,x^3,x^4,x^5)$
is described by the CHS solution \CallanAT\ associated
with the $k_2$ fivebranes stretched in $(016789)$. Similarly, the
supergravity background in the four dimensional space labeled by
$(x^6,x^7,x^8,x^9)$ is that associated with the fivebranes
stretched in $(012345)$. The fact that the background factorizes
will play an important role in our considerations below.

To present the solution, we define \eqn\defyz{\eqalign{ {\bf
y}=&(x^2,x^3,x^4,x^5)~,\cr {\bf z}=&(x^6,x^7,x^8,x^9)~.\cr }} We
have $k_1$ fivebranes localized at the points ${\bf z}={\bf z}_n$,
$n=1,2,3,\cdots, k_1$, and $k_2$ fivebranes localized at ${\bf
y}={\bf y}_a$, $a=1,2,3,\cdots, k_2$. Every pair of fivebranes
from different sets intersects at a point in $\IR^8$, $({\bf y},
{\bf z})=({\bf y}_a, {\bf z}_n)$.

The geometry corresponding to this brane configuration
has the product form
\eqn\twochs{
\IR^{1,1} \times {\rm CHS}_{\bf y} \times {\rm CHS}_{\bf z}~.}
The supergravity background is
\eqn\nssol{\eqalign{\Phi=&\Phi_1({\bf z}) +\Phi_2({\bf y}),\cr
g_{\mu\nu}=&\eta_{\mu\nu},~~~~~\mu,\nu=0,1~,\cr
g_{\alpha\beta}=&e^{2(\Phi_2-\Phi_2(\infty))}\delta_{\alpha\beta},~~~~~
H_{\alpha\beta\gamma}=-\epsilon_{\alpha\beta\gamma\delta}
\partial^{\delta}\Phi_2,~~~~~\alpha,\beta,\gamma,\delta=2,3,4,5,\cr
g_{pq}=&e^{2(\Phi_1-\Phi_1(\infty))}\delta_{pq},~~~~~
H_{pqr}=-\epsilon_{pqrs}\partial^s\Phi_1,~~~~~p,q,r,s=6,7,8,9,\cr}}
where
\eqn\genharm{\eqalign{ e^{2(\Phi_1-\Phi_1(\infty))}=&
1+\sum_{n=1}^{k_1}{l_s^2\over |{\bf z}-{\bf z}_n|^2},\cr
e^{2(\Phi_2-\Phi_2(\infty))}=& 1+\sum_{a=1}^{k_2}{l_s^2\over |{\bf
y}-{\bf y}_a|^2}~. }}
In the rest of this section we describe some properties of this
background, first for the case of coincident fivebranes, ${\bf
z}_n={\bf y}_a=0$, and then for separated ones.

\subsec{Coincident fivebranes}

This is a particularly symmetric case, in which the
brane configuration preserves $SO(4)_{2345}\times SO(4)_{6789}$.
Defining $u=|{\bf y}|$, $v=|{\bf z}|$ and setting $\alpha'=2$,
the geometry \nssol\ takes the form
\eqn\nsso{\eqalign{&ds^2=-(d{x^0})^2+(d{x^1})^2+f_1(v)(dv^2+v^2
d\Omega_v^2)+f_2(u)(du^2+u^2 d\Omega_u^2),\cr
&e^{2(\Phi-\Phi(\infty))}=f_1(v)f_2(u),~~~~~~~
 f_1(v)=1+\frac{2k_1}{v^2},
 ~~~~~~~f_2(u)=1+\frac{2k_2}{u^2}.\cr}}
There is also a flux of the NS $B$ field through the two
three-spheres. As $u,v\to\infty$, the fivebrane background
\nsso\ approaches flat spacetime \met.

As mentioned above, the intersecting fivebrane geometry
contains two copies of the CHS solution corresponding to a single
stack of fivebranes. It has two throats, one associated with $u$,
the other with $v$. The string coupling becomes large when we go down
either of these throats. Note also that since each of the CHS backgrounds
associated with $\bf y$ and $\bf z$ corresponds to an exact solution
of the classical string equations of motion (\ie\ it is a worldsheet CFT with
central charge $c=6$), the factorization \twochs\ is valid in the full
classical string theory and is not restricted to the supergravity
approximation.

To focus on the physics near the intersection of the two sets of
fivebranes we take the near-horizon limit. This is done in the
standard way \refs{\MaldacenaCG,\ItzhakiDD} by rescaling the variables $v$ and $u$ by
$\exp(\Phi_1(\infty))$ and $\exp(\Phi_2(\infty))$, respectively,
and sending $\exp(\Phi_i(\infty))\to 0$. This means that we are
focusing on the region $u,v\sim g_s l_s$ in the full geometry, in
the limit $g_s\to 0$. The resulting near-horizon geometry is
\eqn\nearhor{\IR^{1,1}\times\IR_{\phi_1}\times SU(2)_{k_1}
\times\IR_{\phi_2}\times SU(2)_{k_2}~.} where
$\phi_1=\sqrt{2k_1}\ln v$ and $\phi_2=\sqrt{2k_2}\ln u$.
$\IR_{\phi_i}$ has a   linear dilaton with
\eqn\imprterm{Q_i=\sqrt{2\over k_i}~.} Hence  the worldsheet
central charge of $\phi_i$ is $c_i=1+3Q_i^2$. The $SU(2)$ WZW
models correspond to the angular three-spheres labeled by
$\Omega_v$ and $\Omega_u$, respectively. The $SO(4)\times SO(4)$
symmetry of the brane configuration is realized as the symmetry
group of the two $SU(2)$ CFT's in \nearhor.

Note that the above discussion is only valid when both $k_1$ and
$k_2$ are larger than one. When $k_i=1$ the throat labeled by
$\phi_i$ is absent and we expect the description of the
intersection to be modified. We will return to this point in the
discussion.

As mentioned in the introduction, the near-horizon geometry
\nearhor\ exhibits an unexpected enhanced symmetry. Performing the
rotation (see figure 3)
\eqn\newcoor{\eqalign{
Q\phi=&Q_1\phi_1+Q_2\phi_2~,\cr Qx^2=&Q_2\phi_1-Q_1\phi_2~,\cr }}
where \eqn\defqq{Q=\sqrt{2\over k}; \qquad {1\over k}={1\over
k_1}+{1\over k_2}~,} we find that \nearhor\ takes the form \mh,
 $$\IR^{2,1}\times \IR_\phi\times SU(2)_{k_1}\times SU(2)_{k_2}~, $$
which is indeed invariant under $ISO(2,1)$. This symmetry
enhancement is surprising since the physical origin of the
$\IR^{1,1}$ factor, which consists of directions along the
fivebranes, and of the $x^2$ direction, which is a combination of
directions transverse to one of the two branes, is completely
different. Apparently, in the vicinity of the intersection, the
two become indistinguishable.

\ifig\tspace{The dashed lines represents constant $\phi$ contours
that are parameterized by $x^2$. The dotted lines are constant
$x^2$ contours that are parameterized by $\phi$.}
{\epsfxsize4.0in\epsfbox{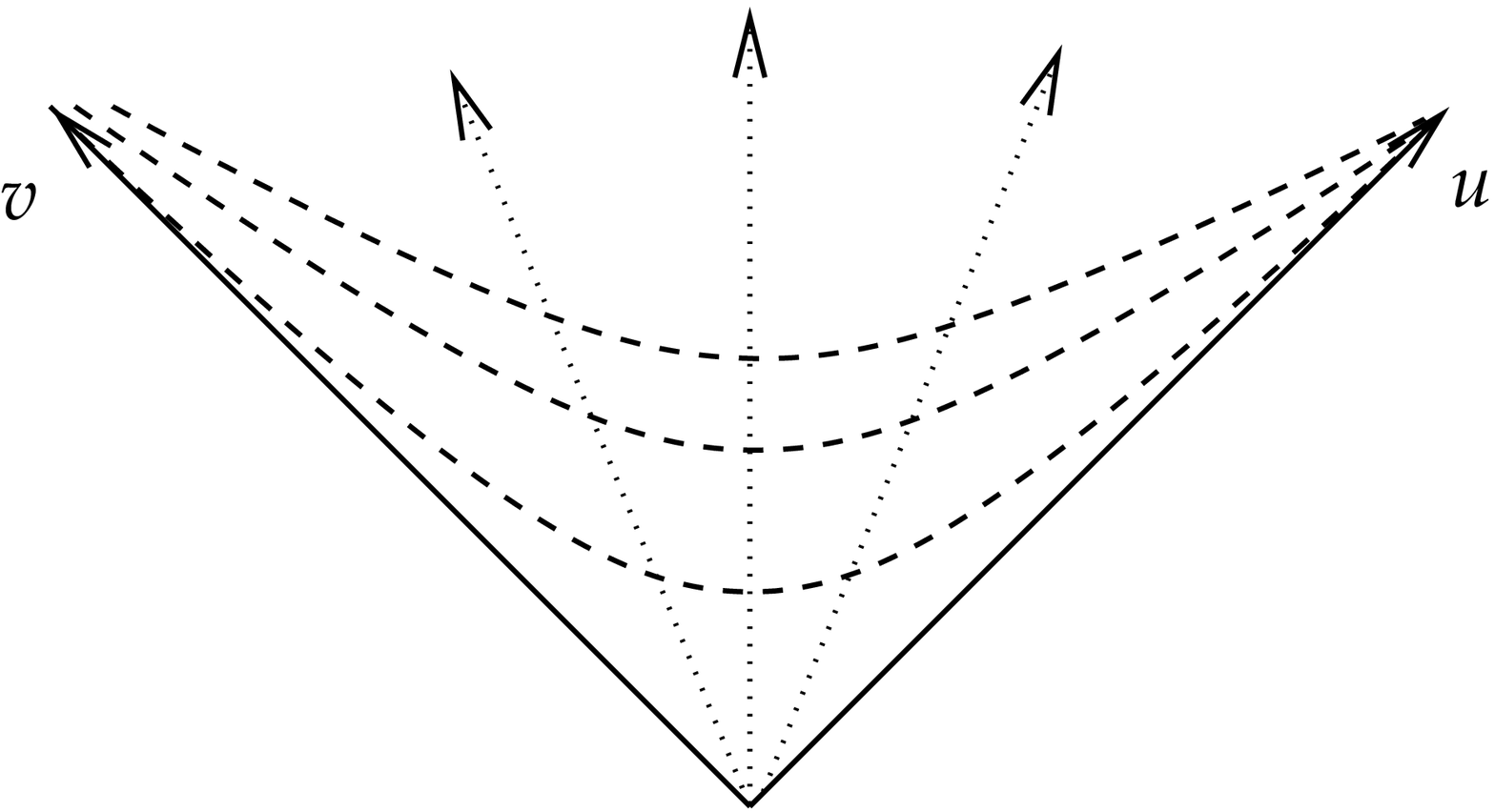}}

The enhanced Poincare symmetry implies also a higher supersymmetry.
Let us briefly recall how this comes about.
Thinking of the near-horizon geometry as a product of an $\IR^{1,1}$
and two CHS geometries, as in \nearhor, one can construct in the usual
way eight spacetime supercharges, by using the worldsheet superpartners
of the coordinates on the spacetime \nearhor. These are ten free fermions:
two fermions $\psi_\pm$ corresponding
to $\IR^{1,1}$, two more that correspond to the two Liouville modes,
$\psi_1$, $\psi_2$, three fermions in the adjoint of $SU(2)_{k_1}$,
$\lambda^{(1)}_a$, $a=3,\pm$, and three in the adjoint of $SU(2)_{k_2}$,
$\lambda^{(2)}_a$. To construct the spacetime supercharges, it is
convenient to bosonize the fermions as follows:
\eqn\bosferm{\eqalign{
\psi_\pm\to & H_1~,\cr
\psi_1,\lambda^{(1)}_3\to & H_2~,\cr
\lambda^{(1)}_\pm\to & H_3~,\cr
\psi_2,\lambda^{(2)}_3\to & H_4~,\cr
\lambda^{(2)}_\pm\to & H_5~,\cr
}}
and similarly for the other worldsheet chirality.

The following spacetime supercharges can be checked to be unbroken
in the background \nearhor:
\eqn\spsup{\eqalign{
Q_+=&\oint{dz\over 2\pi i}
e^{-{\varphi\over2}+{i\over 2}H_1
\pm{i\over 2}(H_2+H_3)
\pm{i\over2}(H_4+H_5)}~,\cr
\bar Q_+=&\oint{d\bar z\over 2\pi i}
e^{-{\bar\varphi\over2}+{i\over 2}\bar H_1
\pm{i\over 2}(\bar H_2-\bar H_3)\pm
{i\over2}(\bar H_4-\bar H_5)}~.\cr
}}
The first line comes from the left-moving sector on the worldsheet,
while the second comes from the right-movers. Note that \spsup\
contains eight supercharges, all with the same chirality in $\IR^{1,1}$.
Note also that the left and right-moving supercharges have opposite
chiralities under $SO(4)_{2345}$ and $SO(4)_{6789}$. This is a well
known property of $NS5$-branes \CallanAT.

The spacetime supercharges \spsup\ have the nice property that
they remain good symmetries when we deform the system by
separating the fivebranes, and when we go beyond the near-horizon
limit and study the full fivebrane geometry \nsso. However, in the
near-horizon geometry \nearhor, we expect to find twice as many
supercharges, due to the enhanced Lorentz symmetry of the
background \mh.

A quick way to construct the extra supercharges is to apply to the
ones we already have, \spsup, the generators of Lorentz
transformations that mix $x^2$ \newcoor\ with $x^\pm$. The
worldsheet superpartner of $x^2$ is $\psi_{x^2}$,
 \eqn\ppssii{Q\psi_{x^2}=Q_2\psi_1-Q_1\psi_2~.}
The Lorentz generators that mix it with $\psi_\pm$ have the form
 \eqn\jtwopm{J_{2\pm}=\oint \psi_{x^2}\psi_\pm+
\oint \bar\psi_{x^2}\bar\psi_\pm~.}
Applying them to the supercharges \spsup\ one finds eight more
supercharges, with the opposite chirality in $\IR^{1,1}$. Since
\jtwopm\ is a symmetry of the background \mh, the resulting
supercharges must be good symmetries as well, a fact that can be
verified directly. Altogether, we find sixteen supercharges which
transform as $(2,2,2)\oplus(2,\bar2,\bar2)$ under $ISO(2,1)\times
SO(4)\times SO(4)$.  It is easy to work out the supersymmetry
algebra of these supercharges.  One unusual aspect of this
superalgebra is that the anticommutator of two supercharges includes
the $SO(4) \times SO(4)$ R-symmetry generators.  This is to be
distinguished from the standard extended supersymmetry algebras
which do not include such generators.  This point was
independently noticed from another point of view in \maldanew.

The near-horizon geometry of intersecting fivebranes \mh,
\nearhor\ can be viewed as a $2+1$ dimensional vacuum of Little
String Theory (LST) \refs{\BerkoozCQ,\SeibergZK}  with sixteen
supercharges. Thus, in studying
it one can use the techniques developed in recent years for
studying such vacua; see \eg\
\refs{\AharonyXN,\AharonyUB\GiveonPX\GiveonTQ-\AharonyVK}. In
particular, one can use the geometry to classify off-shell
observables in the theory, which correspond to non-normalizable
operators, whose wavefunctions are peaked at $\phi\to\infty$. One
can also use it to study bulk physics of delta-function
normalizable modes that live in the throat.

The fact that the string coupling diverges as $\phi\to-\infty$ means
that we cannot use the geometry to study states that are localized there,
or calculate generic correlation functions of the non-normalizable
observables, which receive contributions from the strong coupling region.
However, experience from other vacua of LST suggests that the physics
associated with the strong coupling region can often be at least partially
understood
by studying the low energy field theory on the branes that give rise to
the linear dilaton background.

In our case, that theory was analyzed in section 3, and it is
natural to ask how the picture found there compares to the one
suggested by the geometry. The analysis in section 3.1 led to the
picture summarized in figure 2. The low energy spectrum consists
of three components, \curralg, each of which is located at a
different place. We would like to understand the interpretation of
these modes in the geometry \nsso, and its near-horizon limit \mh.

Consider first the $U(1)$ mode.  In the gauge theory discussion of
section 2 it had support at $u\sim l_s\sqrt{k_2}$ and $v\sim
\l_s\sqrt{k_1}$. From the higher dimensional perspective these two
points are $(u,v)\sim (l_s\sqrt{k_2},0)$ and  $(u,v)\sim
(0,l_s\sqrt{k_1})$.
Below we show that in the strong coupling
limit the support is at $(u,v)\sim (l_s\sqrt{k_2},l_s\sqrt{k_1})$
(see figure 4).

In the geometric description, this mode
corresponds to an excitation of the RR field strength
$F=F_3+F_5+F_7$, of the form
\eqn\selfdual{F\sim dP\wedge\left(H_3^{\bf y}+
\star_{\bf y} H_3^{\bf y} \right)\wedge \left(H_3^{\bf
z}+\star_{\bf z} H_3^{\bf z} \right)~.}
Here $P$ is a chiral scalar field living in $\IR^{1,1}$, $H_3^{\bf
y}$ and $H_3^{\bf z}$ are the field strengths of the NS $B$-field
given in \nssol, and $\star_{\bf y}$ and $\star_{\bf z}$ are the
Hodge duals in the four dimensional spaces labeled by ${\bf y}$
and ${\bf z}$, respectively.

\ifig\tspace{The location of the various currents in the strong
coupling description.} {\epsfxsize3.0in\epsfbox{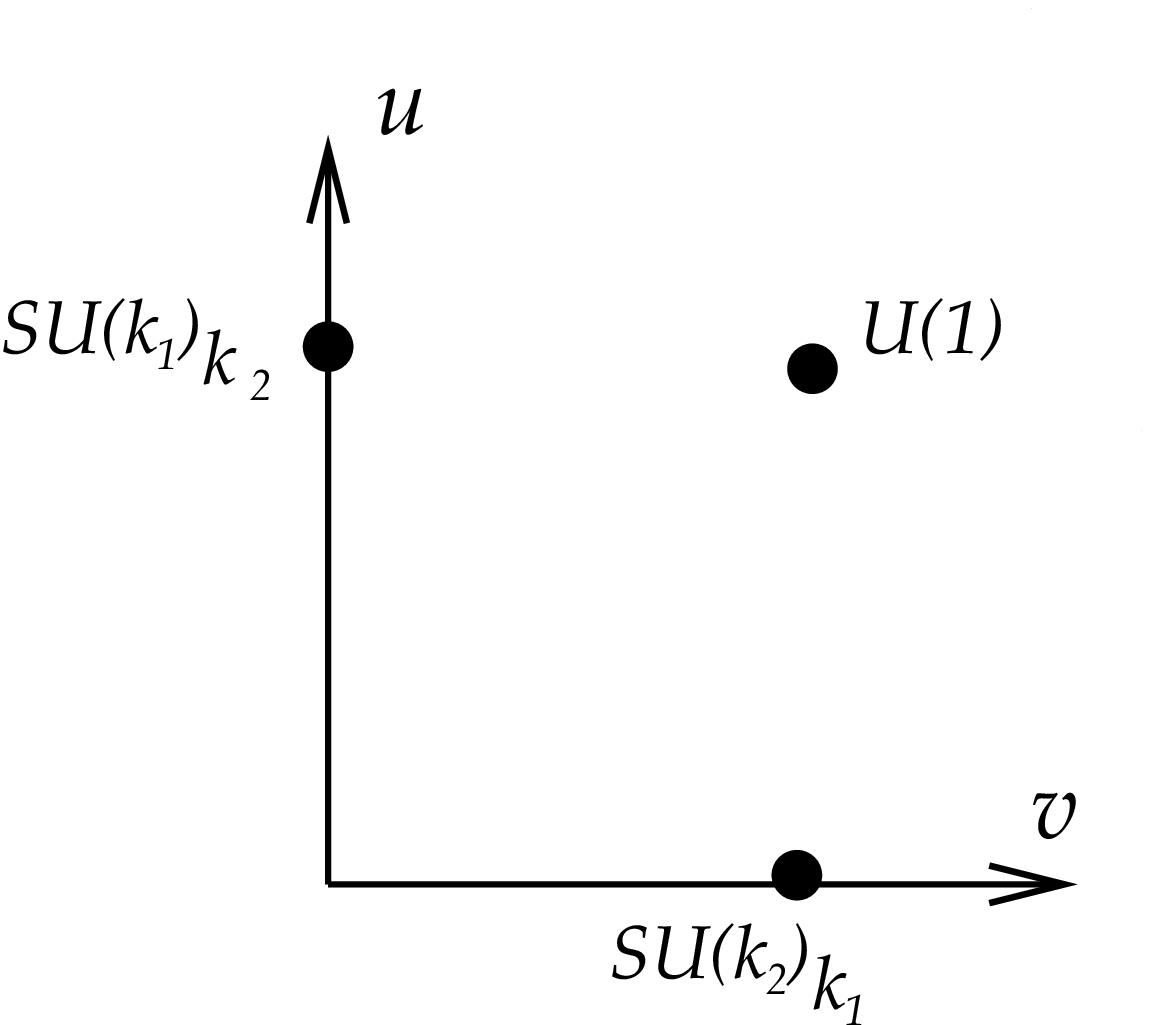}}

Eq. \selfdual\ describes a normalizable mode whose wavefunction is
localized in the transition region between the throats in $u$,
$v$, and the asymptotically flat space far from the fivebranes. At
large $u$, $v$ the five-form part of \selfdual\ falls off like
\eqn\srk{F_5=dP\wedge \left( {1 \over v^3}  dv \wedge
d\Omega_u+ {1 \over u^3} du \wedge d\Omega_v\right) .}
Note that both terms in \srk\ fall off like $u^{-3}v^{-3}$
at large $u$, $v$. For example, in the first term the fall-off
in $v$ is manifest, while the one in $u$ is due to the fact
that the metric on the three-sphere labeled by $\Omega_u$
\nsso\ goes like $ds^2=u^2d\Omega_u^2$ for large $u$.

The behavior \srk\ is in precise agreement with the gauge theory
analysis of sections 2, 3. Indeed, there it was found that the chiral
boson describing the $U(1)$ degree of freedom in \curralg\ corresponds
to gauge field strength $F_{u-}^{(1)}$ on brane 1 that falls off like
$1/u^3$ \solfpm, and field strength on brane 2, $F_{v-}^{(2)}$,
that falls off like $1/v^3$.  To compare to \srk\ one needs to recall
that the gauge fields on the fivebranes correspond to RR operators
in the geometry (see \eg\ \AharonyVK). Thus, the first term in
parenthesis in \srk\ corresponds to ${\rm Tr}F_{v-}^{(2)}$, while
the second corresponds to ${\rm Tr}F_{u-}^{(1)}$. The fact that the
full solution \selfdual\ includes both of these terms is the closed
string analog of the fact that in our gauge theory analysis we
found that the $U(1)$ mode involves both $A^{(1)}$ and $A^{(2)}$.

The left-moving scalar field $P$ \selfdual\ lives outside of the
near-horizon throat. It is a singleton, that does not suffer from
the strong coupling problem of the background \nsso. This agrees
with the gauge theory picture, where this $U(1)$ degree of freedom
is free and decoupled from the non-Abelian dynamics governing the
rest of \curralg.

What about the non-Abelian parts of the current algebra \curralg?
Consider for example the $SU(k_1)_{k_2}$ currents, which according
to figure 4 live at $(u\simeq l_s\sqrt{k_2},v=0)$. From the point
of view of the geometry \nsso\ they correspond to modes of the RR
field strengths which live in the transition region between the
throat and asymptotically flat space in ${\bf y}$, and in the
strong coupling region of the throat in ${\bf z}$. Thus, they
cannot be studied by using supergravity or classical string
theory. From the closed string perspective, the strong coupling
effects are due to the fact that the non-Abelian dynamics that
gives rise to the interacting $SU(k_1)_{k_2}$ WZW model involves
D-branes that descend down the throat labeled by $v$, and become
light there. This should be contrasted with the $U(1)$ current,
which in the gauge theory is free, and in the closed string
description corresponds to a mode that lives outside both throats.

The comparison of the field theory analysis of sections 2, 3 to
the analysis of the geometry of the fivebranes provides us with a
more detailed understanding of the emergence of the enhanced
super-Poincare symmetry of the near-horizon geometry \mh. We see
that in the full problem we have $1+1$ dimensional chiral modes,
which break $ISO(2,1)$, but as we take the near-horizon limit,
they go off to infinity in different directions. The $U(1)$ part
goes to the weakly coupled boundary $\phi\to\infty$, becomes  a
singleton, and decouples. The $SU(k_1)_{k_2}$ an $SU(k_2)_{k_1}$
parts go to  $x^2\to-\infty$ and $x^2\to\infty$, respectively. We
are left with no light $1+1$ dimensional degrees of freedom. The
extreme IR behavior of the theory is described by the Chern-Simons
theory discussed in section 3, which from the closed string point
of view lives in the strong coupling region of \mh.

\subsec{Separated fivebranes}

A well known way to avoid the strong coupling singularity of the
coincident fivebrane background \nsso, is to study the theory
along its Coulomb branch. The near-horizon description is
particularly simple for the case where the two stacks of
fivebranes are placed symmetrically on circles in the respective
transverse $\IR^4$'s. Parameterizing $\left(\IR^4\right)_{6789}$
by the complex coordinates $(a_1, b_1)$ and
$\left(\IR^4\right)_{2345}$ by $(a_2, b_2)$, we place  the first
set of fivebranes at
 \eqn\locone{(a_1^{(n_1)},b_1^{(n_1)})=(0,
 r_1e^{2\pi in_1\over k_1});\;\;\;n_1=1,2,\cdots k_1~,}
and the second set at
  \eqn\loctwo{(a_2^{(n_2)},b_2^{(n_2)})=(0, r_2e^{2\pi
 in_2\over k_2});\;\;\;n_2=1,2,\cdots k_2~.}
Here $r_j$ are the radii of the two circles on which the
fivebranes are located. In the gauge theory on the fivebranes,
$(a_j,b_j)$ correspond to scalar fields in the adjoint of
$U(k_j)$, $(A_j, B_j)$.  The configuration \locone, \loctwo\
corresponds to a point in the Coulomb branch of the gauge theory
at which \eqn\vevbb{\langle B_j\rangle=r_j {\rm diag}( e^{2\pi
i\over k_j},e^{4\pi i\over k_j},\cdots, e^{2\pi i(k_j-1)\over
k_j},1)~.} The masses of the broken gauge bosons are set by the
radii of the circles, $M_W^{(j)}=r_jM_s^2/g_s$.

The background \vevbb\ is a simple generalization of those studied
for a single stack of fivebranes in
\refs{\AharonyXN,\GiveonPX\GiveonTQ-\AharonyVK}, and we can use
the results of these papers in analyzing it. When $M_W^{(j)}\gg
M_s$, the string background corresponding to the configuration
\vevbb\ is perturbative. The separation of the fivebranes
eliminates the strong coupling singularity at the origin, and the
string coupling remains small everywhere. The near-horizon
geometry \nearhor\ is replaced after the deformation by
\eqn\reggeom{\IR^{1,1}\times \left({SL(2)_{k_1}\over
U(1)}\times{SU(2)_{k_1}\over U(1)}\right)/Z_{k_1}\times
\left({SL(2)_{k_2}\over U(1)}\times{SU(2)_{k_2}\over
U(1)}\right)/Z_{k_2}. }
It involves two cigar CFT's, describing the two regulated throats
of \nearhor. The string coupling in the background \reggeom\ is
bounded from above by $g_s^2=M_s^2/ (M_W^{(1)}M_W^{(2)})$,
and therefore is small everywhere. It attains its maximum value at
the tip of both cigars.

Since in this case the fivebrane background is weakly coupled
everywhere, we should be able to see all $k_1k_2$ chiral currents
using perturbative string theory. We next show that this is indeed
the case.

We would like to compute the spectrum of massless excitations in
$1+1$ dimensions in the background \reggeom. Using the techniques
of \refs{\AharonyXN,\GiveonPX\GiveonTQ-\AharonyVK} it can be
checked that the only sector that can contain such states is the
RR sector. All the other sectors have a finite mass gap. Since the
background breaks up into a product of three factors,  the problem
of finding massless modes breaks up into the problem of finding
normalizable zero modes in the different factors. A detailed
discussion of this problem appears in appendix A. Here we only
summarize the results.

Due to the product structure of \reggeom, it is enough to focus on
one of the
 \eqn\ssllssuu{\left( {SL(2)_{k}\over U(1)}\times{SU(2)_{k}\over U(1)}\right) /Z_{k}}
factors. The spectrum of massless RR states on \ssllssuu\ is known
since it plays an important role in analyzing the six dimensional
system of $k$ IIB fivebranes on a circle. In that case, the
massless RR operators correspond holographically to the operators
 \eqn\loweng{{\rm Tr} F_{\mu\nu} B^{2j+1}}
in the low energy $SU(k)$ gauge theory \AharonyVK. When $B$ has an
expectation value of a form analogous to \vevbb, the operators
\loweng\ create states associated with combinations of the
unbroken $U(1)^{k-1}$ gauge bosons
 \eqn\normphot{|O_j\rangle=\sum_{n=1}^k e^{2\pi i n(2j+1)\over k}
F^{(n)}_{\mu\nu}|0\rangle~.} In the near-horizon geometry
\ssllssuu, these states correspond to normalizable wavefunctions,
$O_j$, that behave near the boundary $\phi\to\infty$ like
$\exp(-Qj\phi)$, with $j$ running over the same range as in
\rangejj. If we include the asymptotically flat part of the space
and denote the asymptotic radial direction by $u$, one can show
that for large $u$ (far outside the fivebrane throat) the
wavefunctions $O_j$ behave like \eqn\oojj{O_j\simeq {1\over
u^{2j+4}}~.} Of course, in addition to the $k-1$ wavefunctions
$O_j$ which correspond to the Cartan subalgebra of $SU(k)$, the
full fivebrane geometry includes the wavefunction $O_{-\half}$,
which corresponds to the overall $U(1)$ gauge boson \normphot. As
mentioned earlier, this wavefunction lives in the transition
region between the throat and the flat space far from the branes,
and does not correspond to a normalizable mode in the near-horizon
geometry. Altogether we have $k$ normalizable wavefunctions living
in the full fivebrane geometry, of which $k-1$ live in the throat
\ssllssuu, and one in the transition region.

Given the above spectrum of massless RR modes in the throat of a
single stack of fivebranes, it is not difficult to find the
corresponding spectrum in the geometry of the intersecting ones
\reggeom. We simply take a product of any of the $k_1$
wavefunctions which live in the throat of the first stack of
fivebranes, with any of the $k_2$ wavefunctions which live in the
throat of the second stack. We find $k_1k_2$ chiral massless modes
(see appendix A). These modes naturally split into the following
groups:
 \item{(1)} One mode supported in the transition region
 of both throats in \reggeom.
 \item{(2)} $k_1-1$ modes supported in
 the transition region of throat 2, and deep inside throat 1.
 \item{(3)} $k_2-1$ modes supported in the transition region of
 throat 1, and deep inside throat 2.
  \item{(4)} $(k_1-1)(k_2-1)$ modes supported deep in both throats
  1, 2.

\noindent Only the last class corresponds to normalizable modes in
the near-horizon geometry \reggeom.  In analogy with \loweng, they
can be thought of as states created by the operators
\eqn\quannum{{\rm Tr}\Psi B_1^{2j_1+1}\Psi^* B_2^{2j_2+1},}
acting on the vacuum.

Note that the set of modes found in the geometry is in nice
correspondence with those seen in the gauge theory analysis of
section 3. In particular, the first three types of states
described geometrically above are precisely the states \uonecur,
while the fourth one corresponds to \basrest, a fact that can be
seen directly by substituting \vevbb\ into \quannum. As a check,
the fall-off of the vertex operators of these states agrees with
the gauge theory results \decayuone\ -- \cosss.

\newsec{The big picture}

In sections 2 -- 4 we analyzed some aspects of the dynamics of the
system of intersecting fivebranes from the gauge theory and gravity
points of view. In this section we would like to put the two pictures
together.

We start with the coincident fivebrane configuration. The low
energy modes are spread out in the $(u,v)$ plane in the way
depicted in figure 4.  Now, imagine separating the first stack of
$k_1$ fivebranes as in \locone, while keeping those in the second
stack coincident. In the gauge theory, this means that
$M_W^{(1)}\not=0$, while $M_W^{(2)}=0$. We would like to describe
what happens to the different modes in figure 2 as we increase
$M_W^{(1)}$.

There are a number of different regimes to consider:
 \item{(1)} $M_W^{(1)}\ll {1\over l_s}$.
 \item{(2)} $M_W^{(1)}\gg {1\over l_s}, \;\;\;\;\;
 M_W^{(1)}g_sl_s^2=r_1\ll l_s$.
 \item{(3)} $M_W^{(1)}\gg {1\over l_s},  \;\;\;\;\;
 M_W^{(1)}g_sl_s^2=r_1\gg l_s$.

\noindent Note that while regimes (1) and (2) can be studied in
the near-horizon geometry, regime (3) corresponds to distances
outside the near-horizon limit.

In regime (1), the picture is very similar to that seen for
$M_W^{(1)}=0$ in section 3. Since we did not separate the $k_2$
fivebranes of the second type, the $SU(k_2)_{k_1}$ factor in
figure 4 remains intact. On the other hand, the $SU(k_1)_{k_2}$
factor that lives on the $u$ axis in figure 4 splits into the
unbroken $U(1)^{k_1-1}$, which remains where it was, and the coset
$SU(k_1)_{k_2}/U(1)^{k_1-1}$, which starts moving down in figure
4, in the direction of decreasing $u$.

Regime (2) can be partially analyzed using the closed string
picture. The CHS throat labeled by $v$ is now cut-off, and is
replaced by a cigar, in which the string coupling remains small
throughout. This allows one to use the closed string description
to construct the modes corresponding to $U(1)^{k_1-1}$ currents
that live near the $u$ axis. These are modes of  RR fields of a
form analogous to \selfdual, \eqn\rrff{F\sim
dP\wedge\left(H_3^{\bf y}+ \star_{\bf y} H_3^{\bf y} \right)\wedge
\Lambda_i; \qquad i=1,2,\cdots k_1-1~,} where $\Lambda_i$ are the
normalizable modes in the cigar that entered the discussion of
section 4.2. There, these modes were discussed for the case that
$M_W^{(2)}$ is large, but clearly they exist for $M_W^{(2)}=0$ as
well, since they are localized outside the strong coupling throat
region of the fivebranes of the second type. The modes $\Lambda_i$
are localized at $v\sim l_sg_s$, so the $U(1)^{k_1-1}$ currents
\rrff\ live very close to the $u$ axis, as in regime (1). They are
not significantly influenced by the separation of the $k_1$
fivebranes of the first type.

The remaining part of the $SU(k_1)_{k_2}$ is the coset
$SU(k_1)_{k_2}/U(1)^{k_1-1}$. For $M_W^{(1)}\ll M_s$ it is located
deep in the throat labeled by $v$, but outside the $u$ throat.
For $M_W^{(1)}\gg M_s$ this can no longer be the case, since the
coupling in the $v$ throat is small everywhere. Thus, if the coset
degrees of freedom live outside the $u$ throat, they should be
visible to a perturbative analysis. This is a problem for two
separate reasons. First, we have just discussed the spectrum of
massless states living near the tip of the $v$ cigar and outside
the $u$ throat, and that spectrum does not include the coset
degrees of freedom.  Second, the coset is an interacting CFT, and
we do not expect to be able to see it in a weakly coupled string
description.

Thus, it must be that as $M_W^{(1)}$ increases, the coset $SU(k_1)_{k_2}/U(1)^{k_1-1}$
moves down the $u$ throat, such that by the time we reach regime 2, it is
located deep inside the strongly coupled region  $u\ll l_s$.

In regime 3 the distances between any two fivebranes of the first
type are large, and so we have $k_1$ decoupled systems, each of
which has one fivebrane of the first type intersecting $k_2$
fivebranes of the second type. In terms of figure 2,  roughly
speaking\foot{To obtain a more accurate representation of this
regime, we must take into account the angular three-spheres as
well.} we have a set of  $SU(k_2)_1$ CFT's that are widely
separated in the $v$ direction, by distances of order the
separations  between the fivebranes of the first type.

As one increases $M_W^{(1)}$ from regime (2) to regime (3), the
following must happen. The coset $SU(k_1)_{k_2}/U(1)^{k_1-1}$,
which in regime 2 is located at small $u$, $v$, moves towards
larger $v$ and combine with the $SU(k_2)_{k_1}$ to form the CFT
$\left(SU(k_2)_1\right)^{k_1}$. One can check that the central
charge is precisely right for this, $c_l=k_1(k_2-1)$. In addition,
in regime (3) we have $U(1)^{k_1}$ currents living at $u\sim l_s$,
as before.

So far we described the situation in the case where
$M_W^{(1)}\not=0$, but $M_W^{(2)}=0$. It is not difficult to
generalize the discussion to the case where both are non-zero.
What happened before to $SU(k_1)_{k_2}$ will now happen to both
factors. The $U(1)^{k_1-1}$ factor stays on the $u$ axis, at
$u\sim \sqrt{k_2}l_s$; the $U(1)^{k_2-1}$ factor stays on the $v$
axis at $v\sim \sqrt{k_1}l_s$. The two cosets,
$SU(k_1)_{k_2}/U(1)^{k_1-1}$ and $SU(k_2)_{k_1}/U(1)^{k_2-1}$ move
(as we increase $M_W^{(i)}$), the first towards small $u$, the
second towards small $v$. When we reach the regime $M_W^{(i)}\gg
M_s$, which as we saw in section 4 has a weakly coupled closed
string description, the two cosets meet and combine to form a free
theory, corresponding to the $(k_1-1)(k_2-1)$ currents \basrest.
Note that the fact that the two interacting coset CFT's give rise
when they combine to a free theory is necessary for consistency of
the spacetime dynamics and the closed string description of it,
since the latter can be sent to arbitrarily weak coupling by
tuning $M_W^{(i)}$.

\newsec{Discussion}

There are a number of possible extensions of the results presented
in this paper. In this section we mention some of them.

\subsec{ Thermodynamics}

One natural set of questions concerns the thermodynamics of the
system of intersecting fivebranes. When all the fivebranes in a
given set are coincident we found that the near-horizon geometry
has the form \mh. As usual for gravitational systems, we expect
the high energy thermodynamics to be dominated by black branes.
For asymptotically linear dilaton spacetime such as \mh, the
relevant black brane is
 \eqn\blhl{{SL(2,\IR)_k\over U(1)}\times\IR^2\times SU(2)_{k_1}
 \times SU(2)_{k_2}~.}
Here the level $k$ is related to the numbers of fivebranes $k_1$,
$k_2$ via \defqq. Note that it satisfies the bound $k\ge1$ for
$k_1,k_2\ge 2$, so it is a normalizable state in this regime (see
\refs{\KarczmarekBW,\GiveonMI} for recent discussions of this bound,
and the high energy thermodynamics associated with \blhl). For
$k_1=k_2=2$ one has $k=1$, which is the borderline case for
normalizability.

The background \blhl\ describes a non-extremal solution, in which
a finite energy density is added to the fivebrane intersection.
When the energy density in string units is very large, the
(Euclidean) solution is weakly coupled everywhere, and can be
studied using perturbative string techniques. It is interesting
that this solution preserves the Euclidean group $ISO(2)$; in
particular, it is translationally invariant in $x^2$, and is
invariant under rotations that mix $x^1$ and $x^2$. This provides
further evidence for the fact that the near-horizon dynamics of
coincident fivebranes preserves $ISO(2,1)$. We have seen earlier
that the low energy dynamics is consistent with this assertion,
and now we see that the high energy thermodynamics is
consistent with it as well.

The Bekenstein-Hawking entropy associated with the black hole
\blhl\ has the usual Hagedorn form
\eqn\bhent{S_{\rm bh}=2\pi l_s\sqrt{k}E~.}
This can be compared with the Hagedorn growth of each
set of fivebranes, which has a form similar to \bhent, with
$k\to k_i$. The relation \defqq\ implies that $k<k_1,k_2$, so
that the Hagedorn growth associated with the intersection is
smaller than that associated with each group of fivebranes separately.
This had to be the case, since otherwise the system of non-extremal
parallel fivebranes would develop such intersections dynamically, to
increase its entropy.

\subsec{$k_1=1$}

As mentioned in section 4, our discussion of the near-horizon
geometry of the intersecting fivebrane system is valid when
each stack consists of two or more branes. Indeed, the geometry
\mh, \nearhor\ does not make sense when one of the $k_i$ is equal
to one, since it contains a negative level bosonic $SU(2)$
WZW model.

A very similar problem occurs for the case of the near-horizon geometry
of a single $NS5$-brane, where the resolution is that a single
fivebrane does not develop a linear dilaton throat. It is natural
to ask whether this is also the case for the intersection of a single
fivebrane $(k_1=1)$ with $k_2$ orthogonal ones.

One way to address this questions is the following. T dualizing in
a direction transverse to the $k_2$ fivebranes (and thus along the
single fivebrane orthogonal to them) gives rise to a single $NS5$-brane
in IIA string theory, which wraps an $A_{k_2-1}$ ALE surface
described by the equation $z_1^{k_2} + z_2^2 +z_3^2=0$ in $C^3$.
This system is believed \GiveonZM\ to have a linear dilaton throat
of the form
\eqn\lindilthr{
\IR^{1,1}\times \IR_{\phi} \times S^1 \times {\cal M}_{k_2}/\Gamma,
}
where ${\cal M}_{k_2}$ is the A-series $N=2$ minimal model with
$c=3-6/k_2$ and $\Gamma$ a discrete group.

The geometry \lindilthr\ has very different properties from the
throats discussed in section 4. The Poincare symmetry is not
enhanced beyond $ISO(1,1)$ in this case. This is natural from the
intersecting fivebrane perspective since, as explained in section
4, the $2+1$ dimensional nature of the system for $k_1,k_2>1$
relied on the existence of both throats.

The background \lindilthr\ has some additional properties which
are puzzling from the point of view of the fivebrane
configuration. The near-horizon geometry of the intersecting
fivebranes is expected to have an $SO(4)\times SO(4)$ global
symmetry. The background \lindilthr\ does not have such a symmetry.
Furthermore, out of the eight supercharges that one expects from
the brane perspective, only four are realized linearly.\foot{The
background \lindilthr\ is a special case of the non-critical
superstring construction of \KutasovUA, where the symmetry
structure was analyzed.}

We see that there is some tension between the properties of the
throat geometry \lindilthr\ and those expected from the
near-horizon geometry of the intersecting fivebranes. It would be
interesting to understand the relation between the two better.

\subsec{ Type IIA}

The intersecting $NS5$-brane background \nssol\ is an exact solution
of the classical equations of motion of IIA string theory as
well.\foot{A quick way to see that is to note that the CHS geometry
is an exact worldsheet CFT with $c=6$, and can be used to construct
IIA backgrounds as well.} It is natural to ask how the results
of this paper generalize to that case. Many aspects of the discussion
carry over to the IIA case. In particular, it is still true that the
near-horizon geometry for coincident fivebranes is \mh, and that it
exhibits an enhanced super-Poincare symmetry.

One can also repeat the calculations of section 4.2 and find the spectrum
of massless modes living near the intersection of the branes for $M_W\gg M_s$.
As we show in appendix A, the result is very similar to the IIB case.
Instead of $k_1k_2$ currents \jjiiaa\ associated with left-moving
chiral fermions $\Psi_n^a$, one finds $k_1k_2$ modes of the RR fields that
correspond to products of left and right-moving real (Majorana) fermions,
$\bar\Psi^n_a\Psi_n^a$. It is thus natural to conjecture that each
intersection of two $NS5$-branes carries a Majorana fermion.

Some other elements of the IIB discussion do not have an obvious analog
in the IIA case. In particular, there is no D-brane picture which allows
one to analyze the low energy dynamics associated with the intersection.
One expects that the strong coupling region should be treated
by lifting to eleven dimensions. It is natural to ask whether the full
theory preserves the enhanced symmetry of the near-horizon background
\mh. This requires a better understanding of the fate of the fermions,
and will be left for future work.

\subsec{ Fundamental strings and $AdS_3\times\IR\times S^3 \times
S^3$}

Adding fundamental strings stretched in the directions (01) to the
intersecting $NS5$-branes, and taking the near-horizon limit, one
finds the geometry $AdS_3\times\IR\times S^3 \times S^3$, which
can be thought of as a limit of $AdS_3 \times S^1 \times S^3
\times S^3$ with the radius of the $S^1$ going to infinity. This
system was studied from the point of view of holography in
\refs{\ElitzurMM\deBoerRH-\GukovYM}, but it remains enigmatic.
Some of our results are directly relevant to the study of this
system. For example, the fact that when we take the near-horizon
limit of the fivebranes, the fermions at the intersection
decouple, implies that the same is true after adding the
fundamental strings and taking their near-horizon limit. Also, the
fact that before adding the strings the system has an enhanced
Poincare symmetry should have important implications for the
structure of the spacetime CFT after adding the strings. We will
leave a more detailed discussion of this interesting system for
future work.

\bigskip
\centerline{\bf Acknowledgements} We thank O. Aharony, J. Harvey
and  J. Maldacena for discussions. DK  thanks the Weizmann
Institute and Rutgers NHETC for hospitality during part of this
work. NI thanks the Weizmann Institute and the Enrico Fermi
Institute at the University of Chicago for hospitality. The work
of DK  is supported in part by DOE grant DE-FG02-90ER40560; that
of NS by DOE grant DE-FG02-90ER40542. NI is partially supported by
the National Science Foundation under Grant No.\ PHY 9802484.
Any opinions, findings, and conclusions or recommendations expressed
in this material are those of the authors and do not necessarily
reflect the views of the National Science Foundation.

\appendix{A}{A CFT analysis of separated fivebranes}

\subsec{Conventions}

Consider $N=1$ SCFT on $SU(2)_k$. Denote the bosonic $SU(2)$
currents by $K^a(z)$, $a=1,2,3$, and the fermions by $\chi^a$. The
superconformal generator is \eqn\gsutwo{
G=Q\left({1\over\sqrt2}K^+\chi^-+{1\over\sqrt2}K^-\chi^+
+K^3\chi^3+\chi^+\chi^-\chi^3\right)~.} The currents satisfy the
OPE algebra \eqn\curope{\eqalign{ K^3(z)K^3(w)\sim
&{\half(k-2)\over (z-w)^2}~,\cr K^+(z)K^-(w)\sim &{k-2\over
(z-w)^2}+{2K^3(w)\over z-w}~,\cr K^3(z)K^\pm(w)\sim
&\pm{K^\pm(w)\over z-w}~,\cr }} and the fermions $\chi^a$ satisfy
\eqn\fermsutwo{\eqalign{ \chi^+(z)\chi^-(w)\sim &{1\over z-w}~,\cr
\chi^3(z)\chi^3(w)\sim &{1\over z-w}~.\cr }}

A similar set of conventions will be used for $SL(2)_k$. The
bosonic currents will be denoted by $J^a(z)$, and the fermions by
$\psi^a(z)$. They satisfy the OPE algebra \eqn\slope{\eqalign{
J^3(z)J^3(w)\sim &-{\half(k+2)\over (z-w)^2}~,\cr J^+(z)J^-(w)\sim
&{k+2\over (z-w)^2}-{2J^3(w)\over z-w}~,\cr J^3(z)J^\pm(w)\sim
&\pm{J^\pm(w)\over z-w}~,\cr }} and \eqn\fermsutwo{\eqalign{
\psi^+(z)\psi^-(w)\sim &{1\over z-w}~,\cr \psi^3(z)\psi^3(w)\sim
&-{1\over z-w}~.\cr }} The $N=1$ supercurrent is \eqn\gsl{ G\sim
Q\left({1\over\sqrt2}J^+\psi^-+{1\over\sqrt2}J^-\psi^+
-J^3\psi^3-\psi^+\psi^-\psi^3\right)~.}

\subsec{Six dimensional background -- fivebranes on a circle}

$k$ type IIB fivebranes on a circle are described by the
background
\eqn\circl{\IR^{5,1}\times\left({SL(2,\IR)\over
U(1)}\times {SU(2)\over U(1)}\right)/Z_k~.}
The background must have $(1,1)$ supersymmetry in $5+1$ dimensions.
A technically useful way to think about this system is the following.
Consider the spacetime
\eqn\twelved{\IR^{5,1}\times SL(2)_k\times SU(2)_k~.}
This is a twelve dimensional background with signature $(10,2)$. In order to
get the physical background \circl, one can gauge the null $U(1)$
super-Kac-Moody algebra
\eqn\nulluone{{1\over\sqrt2}(\chi^3-\psi^3)+\theta{Q\over\sqrt2}
\left(J_3^{\rm (tot)}-K_3^{\rm (tot)}\right)~,}
where $J_3^{\rm (tot)}=J_3+\psi^+\psi^-$ is the total $J_3$
current, which receives contributions from bosons and fermions
on the worldsheet, and similarly for $K_3^{\rm (tot)}$.

The gauging of the
null superfield \nulluone\ implies the presence of a bosonic
$(\beta',\gamma')$ ghost system with spin $\half$, and BRST
charge
\eqn\qbrst{Q_{BRST}=\cdots+\oint{dz\over2\pi i}\gamma'(z)
{1\over\sqrt2}(\chi^3-\psi^3)~.}
As usual, we can write
\eqn\bosbg{\gamma'=\eta'e^{\varphi'}~,} where $\varphi'$ is a
canonically normalized scalar field with no linear dilaton.

We can bosonize the free fermions corresponding to \twelved\ as
follows:
\eqn\bosferm{\eqalign{ &\psi^\mu\to H_1,H_2,H_3~,\cr
&\chi^+\chi^-=i\partial H_4~,\cr
&\psi^+\psi^-=i\partial H_5~,\cr
&{1\over\sqrt2}(\chi^3-\psi^3)=e^{iH'}~,\cr }}
such that
\eqn\qqbb{Q_{BRST}=\cdots+\oint{dz\over2\pi i}\eta'
e^{\varphi'+iH'}~.}
The spacetime supercharges take the form
\eqn\spstsup{Q_\alpha^\pm=\oint{dz\over2\pi i}
e^{-{\varphi'\over2}-{i\over2}H'}e^{-{\varphi\over2}} S_\alpha
e^{\pm{i\over2}(H_4+H_5)}~.}
Here $S_\alpha$ is a spinor of ${\rm Spin}(5,1)$ with a
particular chirality, \ie\ $\alpha\in {\bf 4}$.
It is instructive to check that \spstsup\ is indeed BRST
invariant, w.r.t. the BRST charge of the $N=1$ string.

The supercharges $Q_\alpha^\pm$ satisfy the conjugation relation
$\left(Q^+\right)^\dagger=Q^-$, and the anti-commutation relation
$\{Q_\alpha^+,Q_\beta^-\}=\gamma^\mu_{\alpha\beta}P_\mu$. To prove
the latter, one uses the fact that $e^{-\varphi'-iH'}$ is a
picture-changed version of the identity, as can be checked by
applying $\{Q_{BRST},\xi'\}$ to it.

In the type IIB theory, the other worldsheet chirality gives rise
to supercharges $\bar Q_{\dot\alpha}$, which have the opposite
chirality under $SO(5,1)$, \eqn\supbar{\bar
Q_{\dot\alpha}^\pm=\oint{d\bar z\over2\pi i}
e^{-{\bar\varphi'\over2}-{i\over2}\bar H'}e^{-{\bar\varphi\over2}}
\bar S_{\dot \alpha} e^{\pm{i\over2}(\bar H_4+\bar H_5)}~.}
In the IIA case $\bar Q$ has the same chirality as $Q$.

We next discuss some observables and states in the theory from the
present perspective. Consider first the (NS,NS) sector. A class of
non-normalizable vertex operators which correspond to symmetric
traceless operators in the low energy gauge theory is given by
\eqn\ppp{{\rm Tr}\Phi^{i_1}\Phi^{i_2}\cdots\Phi^{i_{2j+2}}(p_\mu)
\leftrightarrow e^{-\varphi-\bar\varphi}\left(\chi\bar\chi
V^{\rm(su)}_j\right)_{j+1;m,\bar m} V^{\rm (sl)}_{j';m,\bar
m}e^{ip\cdot x}~.} Here $\Phi^i$ are scalars in the adjoint of
$SU(k)$. The operators on the l.h.s.\ of \ppp\ are symmetric and
traceless in the indices $(i_1,i_2,\cdots, i_{2j+2})$. On the
r.h.s., the indices $(m,\bar m)$ are the same for $SU(2)$ and
$SL(2)$, due to the constraint that \nulluone\ should vanish. The
$SL(2)$ quantum number $j'$ is determined via the mass-shell
condition \eqn\mmaass{{j(j+1)\over k}-{j'(j'+1)\over k}+\half
p^2=0~.} Note that the $SO(4)$ symmetry corresponding to rotations
of the space transverse to the fivebranes is realized on the
r.h.s. as the $SU(2)_L\times SU(2)_R$ symmetry generated by
$K_a^{\rm (tot)}$, $\bar K_a^{\rm (tot)}$. The gauging of
\nulluone\ breaks this symmetry down to $U(1)\times Z_k$. This is
in agreement with the spacetime picture, where the breaking is due
to the expectation value of one of the complex scalars $\Phi^i$,
which we will denote by $B$ \refs{\GiveonPX,\GiveonTQ}.

We are particularly interested in the spectrum of on-shell
massless states. These correspond to principal discrete series
states in $SL(2,\IR)$, with $m=\bar m=j'+1$ in \ppp. The
masslessness condition implies, via \mmaass\ that $j'=j$. Hence we
have normalizable states of the form \ppp\ with $m=\bar
m=j+1=j'+1$, or more explicitly,
\eqn\onsb{e^{-\varphi-\bar\varphi}\chi^+\bar\chi^+V_{j;j,j}^{\rm
(su)} V_{j;j+1,j+1}^{\rm (sl)}e^{ip\cdot x}}
with $p^2=0$. These
correspond in the low energy gauge theory to the states
\eqn\lowee{{1\over 2j+2}{\rm Tr} B^{2j+2}|0\rangle= \sum_{l=1}^k
b_l e^{2\pi i{l(2j+1)\over k}}|0\rangle~,} with $b_l$ the
eigenvalues of $B$ and $2j+1=1,2,3,\cdots, k-1$. Thus, we see here
the $k-1$ massless fields corresponding to the eigenvalues of $B$
in the adjoint of $SU(k)$.

We next move on to the (R,R) sector, in particular the operators
dual to ${\rm Tr} F_{\mu\nu}B^{2j+1}$. The normalizable states
corresponding to these operators have the form
\eqn\vertgauge{e^{-{\varphi\over2}-{\bar\varphi\over2}-
{\varphi'\over2}-{i\over2}H'-{\bar\varphi'\over2}-{i\over2}\bar
H'}
\xi_{\mu\nu}\gamma^{\mu\nu}_{\dot\alpha\alpha}S_{\dot\alpha}\bar
S_\alpha e^{ip\cdot x} e^{{i\over2}(H_4+\bar H_4)}V^{\rm
(su)}_{j;j,j} e^{-{i\over2}(H_5+\bar H_5)}V^{\rm (sl)}_{j;j+1,j+1}~.
}
Note that this satisfies $J_3^{\rm(tot)}=K_3^{\rm (tot)}$ and
$\bar J_3^{\rm(tot)}=\bar K_3^{\rm (tot)}$, as necessary for
\nulluone\ (of course, the same is true for \onsb). \vertgauge\
are normalizable vertex operators corresponding to the
$U(1)^{k-1}$ photons
\eqn\nouone{\sum_{l=1}^kF_{\mu\nu}^{(l)}e^{2\pi i{l(2j+1)\over
k}}~.} Again, as in \lowee, the overall $U(1)$ degree of freedom is
absent. The states \vertgauge, \nouone\ were discussed in \AharonyVK;
the description here gives a particularly simple construction
of these states.

\subsec{Two dimensional background -- intersecting fivebranes}

We next move on to the case of two sets of fivebranes in IIB string theory:

\item{(1)} $k_1$ $NS5$-branes stretched in $(012345)$. \item{(2)}
$k_2$ $NS5$-branes stretched in $(016789)$.

We will spread the fivebranes in each set on a circle, as before,
to regularize the problem. The background of interest will now be
\eqn\fourteend{\IR^{1,1}\times SL(2)_{k_1}\times SU(2)_{k_1}
\times SL(2)_{k_2}\times SU(2)_{k_2}~.}
This is a fourteen dimensional spacetime of signature $11+3$. To
get to the ten dimensional fivebrane geometry, we need to gauge
two null $U(1)$'s of the form \nulluone, one for each
$SL(2)\times SU(2)$ factor in \fourteend.

For brevity, we are going to omit below factors analogous to
$e^{-\half(\varphi'+iH')}$ in \spstsup, etc. As we saw in the
previous subsection, these factors play no role in the discussion
since $e^{-r(\varphi'+iH')}$ has dimension zero for all $r$, and
if $r\in Z_+$, the operator is proportional to the identity
operator (after picture changing).

We will use the following notation for the free fermions:
\eqn\fermbos{\eqalign{ \psi_0\psi_1=&i\partial H~,\cr
\chi^{(1)}_+\chi^{(1)}_-=&i\partial H_1~,\cr
\psi^{(1)}_+\psi^{(1)}_-=&i\partial H_2~,\cr
\chi^{(2)}_+\chi^{(2)}_-=&i\partial H_3~,\cr
\psi^{(2)}_+\psi^{(2)}_-=&i\partial H_4~.\cr }}
As we have seen, the
SUSY corresponding to this fivebrane configuration in type IIB is
four supercharges from the left and four from the right, all with
the same chirality in $(01)$. These supercharges can be constructed
in analogy to \spstsup, \supbar:
\eqn\supalg{\eqalign{Q^{\pm,\pm}=&\oint
e^{-{\varphi\over2}+{i\over2}H\pm{i\over2}(H_1+H_2)\pm{i\over2}(H_3+H_4)}~,\cr
\bar Q^{\pm,\pm}=&\oint e^{-{\bar\varphi\over2}+{i\over2}\bar H
\pm{i\over2}(\bar H_1+\bar H_2)\pm{i\over2}(\bar H_3+\bar H_4)}~.\cr
}}
Let us start with the question of what normalizable massless
states there are in this geometry. We need to consider principal
discrete series states in each $SL(2)$. It is not difficult to see
that no analogs of the
(NS,NS) states \onsb\ can survive the GSO projection. The only
states that can possibly make it are (RR) ones. They are described
by the vertex operators (compare to \vertgauge)
\eqn\chirsc{\eqalign{
&e^{-{\varphi\over2}-{\bar\varphi\over2}} e^{{i\over2}(H+\bar
H)}e^{ip\cdot x} e^{{i\over2}(H_1+\bar
H_1)}V^{\rm(su_1)}_{j_1;j_1,j_1} e^{-{i\over2}(H_2+\bar
H_2)}V^{\rm(sl_1)}_{j_1;j_1+1,j_1+1}\cr &e^{{i\over2}(H_3+\bar
H_3)}V^{\rm(su_2)}_{j_2;j_2,j_2} e^{-{i\over2}(H_4+\bar
H_4)}V^{\rm(sl_2)}_{j_2;j_2+1,j_2+1}~.\cr }}
For each $(j_1,j_2)$ with $2j_i=0,1,2,\cdots, k_i-2$, we have a
chiral scalar. In the convention where the supercharges satisfy
$Q^2\sim P^+$, the states \chirsc\ have $P^+=0$, $P^-\not=0$.
The total number of chiral scalars is $(k_1-1)(k_2-1)$. Looking
at the transformation properties under $SO(4)_{2345}\times SO(4)_{6789}$,
we see that these scalars carry quantum numbers of
\eqn\quannum{{\rm Tr}\Psi B_1^{2j_1+1}\Psi^*
B_2^{2j_2+1}~.}
They correspond to the currents \basrest.

A useful interpretation of the $(k_1-1)(k_2-1)$ holomorphic
currents \chirsc\ is as the RR ground states of the CFT on the
product of cigars and minimal models. Each minimal model has
$k_i-1$ chiral operators, which are mapped by spectral flow to
$k_i-1$ RR ground states. At a generic point in the moduli space
of fivebrane positions, all these are normalizable, and we have
$(k_1-1)(k_2-1)$ RR ground states.

It is easy to generalize the above analysis to the case of $NS5$-branes
intersecting in IIA string theory. In the expression for the supercharges
\supalg, and RR vertex operators \chirsc, one simply takes
$H\to H$, $\bar H\to -\bar H$. This leads to a superalgebra with
four left-moving and four right-moving (in spacetime) supercharges,
and the RR states \chirsc\ have the quantum numbers of
\eqn\quantwoa{{\rm Tr}\Psi B_1^{2j_1+1}\bar\Psi
B_2^{2j_2+1}~,}
where $\Psi_n^a$ are real left-moving (Majorana Weyl) fermions, and
$\bar\Psi^n_a$ are their right-moving counterparts.

\listrefs
\end